\documentclass[]{aa}  

\usepackage{graphicx}
\usepackage{txfonts}
\usepackage{lscape}  
\usepackage{color}
\usepackage[colorlinks,citecolor=blue]{hyperref}

\usepackage{multirow}  
\usepackage{xcolor}

\begin{document}

\title{R-matrix electron-impact excitation data for the N-like iso-electronic sequence}


\author{Junjie Mao \inst{\ref{inst_strath}}
    \and N. R. Badnell \inst{\ref{inst_strath}}
    \and G. Del Zanna \inst{\ref{inst_damtp}}
    }

   \institute{Department of Physics, University of Strathclyde, Glasgow G4 0NG, UK\label{inst_strath}
   \and Department of Applied Mathematics and Theoretical Physics, University of Cambridge, Cambridge CB3 0WA, UK\label{inst_damtp}
   }


 
\abstract
{Spectral lines from N-like ions can be used to measure the temperature and density of various types of astrophysical plasmas. The atomic databases of astrophysical plasma modelling codes still have room for improvement in their electron-impact excitation data sets for N-like ions, especially $R$-matrix data. This is particularly relevant for future observatories (e.g. Arcus) which will host high-resolution spectrometers.}
{We aim to obtain level-resolved effective collision strengths for all transitions up to $nl=5d$ over a wide range of temperatures for N-like ions from \ion{O}{II} to \ion{Zn}{XXIV} (i.e., O$^{+}$ to Zn$^{23+}$) and to assess the accuracy of the present work. We also examine the impact of our new data on plasma diagnostics by modelling solar observations with CHIANTI. }
{We have carried-out systematic $R$-matrix calculations for N-like ions which included 725 fine-structure target levels in both the configuration interaction target and close-coupling collision expansions. The $R$-matrix intermediate coupling frame transformation method was used to calculate the collision strengths, while the AUTOSTRUCTURE code was used for the atomic structures.}
{We compare the present results for selected ions with those in archival databases and the literature. The comparison covers energy levels, oscillator strengths, and effective collision strengths. We show examples of improved plasma diagnostics when compared to CHIANTI models which use only distorted wave data as well as some which use previous $R$-matrix data. The electron-impact excitation data are archived according to the Atomic Data and Analysis Structure (ADAS) data class {\it adf04} and will be available in OPEN-ADAS. The data can be used to improve the atomic databases for astrophysical plasma diagnostics. }
{}
\titlerunning{Electron-impact excitation data for N-like ions}
\authorrunning{J. Mao et al.}       
\maketitle

\section{Introduction}
\label{sct:intro}
Plasma codes widely used in astronomy (e.g., AtomDB\footnote{http://www.atomdb.org/Webguide/webguide.php}, CHIANTI\footnote{https://www.chiantidatabase.org/}, SPEX\footnote{https://www.sron.nl/astrophysics-spex}) aim to have extensive and accurate atomic data for a wide range of ions and processes to enable the spectroscopic diagnosis of various types of astrophysical plasmas. For instance, the outer solar atmosphere \citep[e.g.,][]{dza18a}, planetary nebulae \citep[e.g.,][]{ost06}, and ionized outflows in active galactic nuclei \citep{mao17}. Nevertheless, the latest atomic databases used by astrophysical plasma codes are still not as complete and accurate as we would wish. Improvement in both completeness and accuracy are essential, especially for the next generation of spectrometers to be found aboard future observatories like Arcus \citep{smi16}, ATHENA/X-ray Integral Field Unit \citep{bar18}, and Hot Universe Baryon Surveyor \citep{cui20}. 

Electron-impact excitation is one of the dominant atomic processes to populate excited levels (including the metastable levels), which subsequently leads to emission lines from excited levels to the ground and metastable levels, as well as absorption lines from excited levels. Thus, the precision of plasma diagnostics relies on the accuracy of the electron-impact excitation data.

In terms of $R$-matrix electron-impact excitation data, systematic calculations for many iso-electronic sequences (Li-, Be-, B-, F-, Ne-, Na-, and Mg-like) have been performed over the past decade \citep[see][for a review]{bad16} and \citet[][for the most recent C-like one]{mao20}. When $R$-matrix data are not available, then either interpolated data (from adjacent ions in the same iso-electronic sequence) or less accurate distorted wave data are used, if available. 


Most of the existing $R$-matrix calculations have been performed for individual N-like ions. The number of energy levels of the target ion and the temperature range of the effective collision strength vary significantly. For instance, \citet{tay07} provided effective collision strengths for \ion{O}{II} between 47 energy levels over a temperature range of $10^{3.3-5}$~K. \citet{ram97} obtained  effective collisions strength for \ion{Mg}{VI} between 23 energy levels over a temperature range of $10^{5.0-6.1}$~K. \citet{lia11} provided effective collision strengths for \ion{S}{X} between 84 energy levels over a temperature range of $10^{4.3-8.3}$~K. \citet{wit07} calculated  effective collisions strength for \ion{Fe}{XX} between 302 energy levels over a temperature range of $10^{2.0-8.3}$~K.

On the other hand, \citet{wan18} presented a systematic $R$-matrix calculation for N-like ions, from \ion{Na}{V} to \ion{Ca}{XIV}. 272 energy levels were included for each target ion. Effective collision strengths for transitions from the ground level are available over the temperature range $10^{4.0-7.0}$~K .

Atomic data with a larger number of energy levels would be preferred by observers. With advances in technology, we are able to observe more and more transitions that can be used for plasma diagnostics. A wide temperature range would also be favored by observers probing astrophysical plasmas ranging from the near-infrared band to the X-ray band. For instance, collisional ionized plasmas (up to several million degrees Kelvin), photoionized plasmas exposed to stars or active galactic nuclei, non-equilibrium ionization plasmas (often observed in supernova remnants), and the interface between the hot and cold gas.  

Following systematic intermediate coupling frame transformation (ICFT) $R$-matrix calculations for C-like ions \citep{mao20}, here we present similar calculations for N-like ions from \ion{N}{II} to \ion{Zn}{XXIV} (i.e., N$^{+}$ to Zn$^{23+}$). For each ion, we obtain effective collision strengths between 725 levels over a temperature range spanning five orders of magnitude. 

We describe the structure and collision calculations in Section~\ref{sct:str} and Section~\ref{sct:col}, respectively. Results and discussions are provided in Section~\ref{sct:res} and \ref{sct:dis}, respectively. A summary is provided in Section~\ref{sct:sum}. In addition, we provide a supplementary package at Zenodo\footnote{\href{https://sandbox.zenodo.org/record/661331}{DOI: 0.5072/zenodo.661331}}. This package includes the input files of the structure and collision calculations, atomic data from the present work, archival databases and literature. This package also includes scripts used to create the figures presented in this paper. 

\section{Method}
\label{sct:mo}
We adopted the same approach for the structure and collision calculations, described in Section~\ref{sct:str} and Section~\ref{sct:col}, as detailed in \citet{mao20} for C-like ions. The main difference for N-like ions is that we included a total of 725 fine-structure levels in both the configuration-interaction target expansion and the close-coupling collision expansion. These levels arise from the 27 configurations listed in Table~\ref{tbl:cfg}. 

\begin{table*}
\caption{List of configurations used for the structure and collision calculations. }
\label{tbl:cfg}
\centering
\begin{tabular}{cl|cl|cl}
\hline\hline
\noalign{\smallskip} 
Index & Conf. & Index & Conf. & Index & Conf. \\
\noalign{\smallskip} 
\hline
\noalign{\smallskip} 
1 & $2s^22p^3$ & 2 & $2s2p^4$ & 3 & $2p^5$ \\
\noalign{\smallskip} 
4 & $2s^22p^23s$ & 5 & $2s^22p^23p$ & 6 & $2s^22p^23d$ \\
\noalign{\smallskip} 
7 & $2s2p^33s$ & 8 & $2s2p^33p$ & 9 & $2s2p^33d$ \\
\noalign{\smallskip} 
10 & $2p^43s$ & 11 & $2p^43p$ & 12 & $2p^43d$ \\
\noalign{\smallskip} 
13 & $2s^22p^24s$ & 14 & $2s^22p^24p$ & 15 & $2s^22p^24d$ \\
\noalign{\smallskip} 
16 & $2s^22p^24f$ & 17 & $2s2p^34s$ & 18 & $2s2p^34p$ \\
\noalign{\smallskip} 
19 & $2s2p^34d$ & 20 & $2s2p^34f$ & 21 & $2p^44s$ \\
\noalign{\smallskip} 
22 & $2p^44p$ & 23 & $2p^44d$ & 24 & $2p^44f$ \\
\noalign{\smallskip} 
25 & $2s^22p^25s$ & 26 & $2s^22p^25p$ & 27 & $2s^22p^25d$  \\
\noalign{\smallskip} 
\hline
\end{tabular}
\end{table*}

\subsection{Structure}
\label{sct:str}
We used AUTOSTRUCTURE \citep{bad11} to calculate the target atomic structure. The wave functions are calculated via diagonalizing the Breit-Pauli Hamiltonian \citep{eis74}. The one-body relativistic terms: mass-velocity, nuclear plus Blume \& Watson spin-orbit and Darwin, are included perturbatively. We use the Thomas-Fermi-Dirac-Amaldi model for the electronic potential. The $nl$-dependent scaling parameters \citep{nus78} are obtained following the procedure presented in \citet{mao20} without manual re-adjustment. This ensures that we do not introduce arbitrary changes across the iso-electronic sequence. We list the scaling parameters for the 13 atomic orbitals from $1s$ to $5d$ in Table~\ref{tbl:sca_par}. These scaling parameters are used for both the structure and collision calculations for all the ions ($Z=8-30$) in the sequence. 

As shown later in Section~\ref{sct:dis}, the atomic structure obtained in the present work shows relatively large deviations with respect to experiment for low-charge ions (e.g., \ion{O}{II}, \ion{Mg}{VI}) and  low-lying energy levels. This is because we use a unique set of non-relativistic orthogonal orbitals \citep{ber95} --- this is required by the ICFT $R$-matrix method, the calculations with which are described next (Section~\ref{sct:col}). 

The Dirac $R$-matrix method (DARC) and associated multi-configuration Dirac-Fock (MCDF) structure use a unique set of orthogonal orbitals.
The B-spline $R$-matrix method (BSR) and associated multi-configuration Hartree-Fock (MCHF) structure can use non-unique and/or non-orthogonl orbitals. 
These approaches are more computationally expensive for the scattering calculations, especially the BSR method.
\citet{dza19} performed a detailed case study for \ion{N}{iv} where they generated line intensities from three different available atomic data sets  (AUTOSTRUCTURE + ICFT, MCHF + BSR, MCDF + DARC) which used the same set of target states. They found agreement between all of the spectroscopically relevant line intensities (within 20\%), which provides confidence in the reliability of the present calculations for plasma diagnostics.

\begin{longtab}
\begin{landscape}
\begin{longtable}{lccccccccccccc}
\caption{\label{tbl:sca_par} Thomas-Fermi-Dirac-Amaldi potential scaling parameters used in the AUTOSTRUCTURE calculations for the N-like iso-electronic sequence. $Z$ is the atomic number, e.g., 14 for silicon.}\\
\hline\hline 
\noalign{\smallskip} 
$Z$ & 1s & 2s & 2p & 3s & 3p & 3d & 4s & 4p & 4d & 4f & 5s & 5p & 5d \\ 
\noalign{\smallskip} 
\hline 
\noalign{\smallskip} 
 8 & 1.47243 & 1.19629 & 1.13955 & 1.20168 & 1.16406 & 1.17932 & 1.18660 & 1.14257 & 1.15107 & 1.16400 & 1.18518 & 1.13943 & 1.15198 \\ 
\noalign{\smallskip} 
 9 & 1.45362 & 1.17931 & 1.12446 & 1.22623 & 1.13000 & 1.22212 & 1.20449 & 1.18226 & 1.21296 & 1.26254 & 1.21160 & 1.13470 & 1.23432 \\ 
\noalign{\smallskip} 
10 & 1.44249 & 1.18077 & 1.12217 & 1.23875 & 1.19102 & 1.23133 & 1.22180 & 1.14969 & 1.23266 & 1.33308 & 1.20665 & 1.15497 & 1.20116 \\ 
\noalign{\smallskip} 
11 & 1.43358 & 1.18277 & 1.12101 & 1.23576 & 1.18272 & 1.25018 & 1.23599 & 1.15849 & 1.22453 & 1.42000 & 1.20838 & 1.17514 & 1.21844 \\ 
\noalign{\smallskip} 
12 & 1.42605 & 1.18458 & 1.12034 & 1.24622 & 1.19278 & 1.24875 & 1.22033 & 1.17594 & 1.23092 & 1.35497 & 1.21371 & 1.16930 & 1.22455 \\ 
\noalign{\smallskip} 
13 & 1.41966 & 1.18610 & 1.11999 & 1.24145 & 1.19028 & 1.25897 & 1.21910 & 1.17880 & 1.23582 & 1.29772 & 1.21143 & 1.16123 & 1.24206 \\ 
\noalign{\smallskip} 
14 & 1.41417 & 1.18747 & 1.11982 & 1.24435 & 1.18700 & 1.24794 & 1.22099 & 1.16679 & 1.24421 & 1.31233 & 1.22326 & 1.16993 & 1.22339 \\ 
\noalign{\smallskip} 
15 & 1.40948 & 1.18866 & 1.11978 & 1.24183 & 1.18889 & 1.25273 & 1.21969 & 1.17337 & 1.24281 & 1.17666 & 1.23428 & 1.16560 & 1.23656 \\ 
\noalign{\smallskip} 
16 & 1.40539 & 1.18970 & 1.11982 & 1.24041 & 1.18811 & 1.25768 & 1.21395 & 1.16949 & 1.24177 & 1.08911 & 1.24444 & 1.17470 & 1.22993 \\ 
\noalign{\smallskip} 
17 & 1.40181 & 1.19062 & 1.11991 & 1.24097 & 1.18692 & 1.25579 & 1.22169 & 1.17845 & 1.25148 & 1.17549 & 1.21962 & 1.18065 & 1.23149 \\ 
\noalign{\smallskip} 
18 & 1.39863 & 1.19143 & 1.12003 & 1.24128 & 1.17916 & 1.25473 & 1.21851 & 1.17816 & 1.24284 & 1.18274 & 1.22862 & 1.17825 & 1.23506 \\ 
\noalign{\smallskip} 
19 & 1.39583 & 1.19215 & 1.12018 & 1.24105 & 1.18792 & 1.25370 & 1.23000 & 1.18509 & 1.24640 & 1.19295 & 1.21427 & 1.18596 & 1.23802 \\ 
\noalign{\smallskip} 
20 & 1.39331 & 1.19280 & 1.12034 & 1.24109 & 1.18776 & 1.25550 & 1.22640 & 1.18039 & 1.24187 & 1.21543 & 1.23527 & 1.17978 & 1.23773 \\ 
\noalign{\smallskip} 
21 & 1.39104 & 1.19339 & 1.12050 & 1.24112 & 1.18772 & 1.25599 & 1.23262 & 1.18849 & 1.24157 & 1.23127 & 1.21970 & 1.18175 & 1.24619 \\ 
\noalign{\smallskip} 
22 & 1.38897 & 1.19392 & 1.12067 & 1.24124 & 1.18789 & 1.25571 & 1.23630 & 1.18272 & 1.24200 & 1.23463 & 1.23999 & 1.17833 & 1.24066 \\ 
\noalign{\smallskip} 
23 & 1.38711 & 1.19441 & 1.12084 & 1.24135 & 1.18808 & 1.25577 & 1.23598 & 1.18931 & 1.24350 & 1.25423 & 1.21108 & 1.18644 & 1.24199 \\ 
\noalign{\smallskip} 
24 & 1.38540 & 1.19485 & 1.12101 & 1.24146 & 1.18826 & 1.25581 & 1.23905 & 1.18390 & 1.24864 & 1.24815 & 1.22650 & 1.17750 & 1.24284 \\ 
\noalign{\smallskip} 
25 & 1.38384 & 1.19526 & 1.12116 & 1.24157 & 1.18843 & 1.25584 & 1.23516 & 1.18879 & 1.24349 & 1.26176 & 1.21425 & 1.18148 & 1.24192 \\ 
\noalign{\smallskip} 
26 & 1.38241 & 1.19563 & 1.12132 & 1.24166 & 1.18859 & 1.25586 & 1.23747 & 1.18763 & 1.24493 & 1.25717 & 1.22254 & 1.18043 & 1.24294 \\ 
\noalign{\smallskip} 
27 & 1.38109 & 1.19597 & 1.12147 & 1.24176 & 1.18874 & 1.25587 & 1.23698 & 1.18816 & 1.24499 & 1.26062 & 1.22636 & 1.18296 & 1.24314 \\ 
\noalign{\smallskip} 
28 & 1.37992 & 1.19629 & 1.12162 & 1.24184 & 1.18889 & 1.25587 & 1.23733 & 1.18871 & 1.24526 & 1.26327 & 1.22662 & 1.18409 & 1.24347 \\ 
\noalign{\smallskip} 
29 & 1.37879 & 1.19659 & 1.12176 & 1.24193 & 1.18903 & 1.25587 & 1.23718 & 1.18879 & 1.24544 & 1.26621 & 1.22732 & 1.18521 & 1.24376 \\ 
\noalign{\smallskip} 
30 & 1.37773 & 1.19686 & 1.12190 & 1.24201 & 1.18916 & 1.25586 & 1.23736 & 1.18906 & 1.24565 & 1.26821 & 1.22974 & 1.18635 & 1.24407 \\ 
\noalign{\smallskip} 
\hline 
\end{longtable}
\end{landscape}
\end{longtab}

\subsection{Collision}
\label{sct:col}
The ICFT $R$-matrix collision calculation consists of an energy-independent inner-region and energy-dependent outer-region calculation \citep{bur11} for each ion. For both, we included angular momenta up to $2J=22$ and $2J=76$ for the exchange and non-exchange, respectively, calculations. For higher angular momenta, up to infinity, we used the top-up formula of the Burgess sum rule \citep{bur74} for dipole allowed transitions and a geometric series for the non-dipole allowed transitions \citep{bad01}. 

The energy-dependent outer-region $R$-matrix calculation consists of three separate calculations, for each ion. Firstly, an exchange calculation using a fine energy mesh between the first and last thresholds to sample the resonances. Along the iso-electronic sequence, the number of sampling points in the fine energy mesh was increased with atomic number, ranging from $\sim3600$ for \ion{O}{II} to $\sim30000$ for \ion{Zn}{XXIV}, to strike the balance between the computational cost and resonance sampling. 
Secondly, an exchange calculation using a coarse energy mesh from the last threshold up to three times the ionization potential. We used $\sim1000$ points for all the ions in the iso-electronic sequence for this coarse energy mesh. 
Thirdly, a non-exchange calculation using another coarse energy mesh, this time from the first threshold up to three times the ionization potential. We used $\sim1400$ energy points for all ions in the iso-electronic sequence. Since this coarse energy mesh covers the resonance region, post-processing is necessary to remove any unresolved resonances in the ordinary collision strengths. 

The effective collision strength ($\Upsilon_{ij}$) for electron-impact excitation is obtained by convolving the ordinary collision strength ($\Omega_{ij}$) with the Maxwellian energy distribution:
\begin{equation}
    \Upsilon_{ij} = \int \Omega_{ij}~\exp\left(-\frac{E}{kT}\right)~d\left(\frac{E}{kT}\right)~,
\label{eq:upsilon}
\end{equation}
where $E$ is the kinetic energy of the scattered free electron, $k$ the Boltzmann constant, and $T$ the electron temperature of the plasma. Ordinary collision strengths at high collision energies are required to obtain effective collision strengths at high temperatures. We used AUTOSTRUCTURE to calculate the infinite-energy Born and dipole line strength limits. Between the last calculated energy point and the two limits, we interpolate taking into account the type of transition in the Burgess--Tully scaled domain \citep[i.e., the quadrature of reduced collision strength over reduced energy][]{bur92} to complete the Maxwellian convolution (Equation~\ref{eq:upsilon}).

\section{Results}
\label{sct:res}
We have obtained $R$-matrix electron-impact excitation data for the N-like iso-electronic sequence from \ion{O}{II} to \ion{Zn}{XXIV} (i.e., ${\rm O^+}$ and ${\rm Zn^{23+}}$). Our effective collision strengths cover five orders of magnitude in temperature $(z+1)^2(2\times10^1,~2\times10^6)~{\rm K}$, where $z$ is the ionic charge (e.g., $z=7$ for \ion{Si}{VIII}). 

The effective collision strength data will be archived according to the Atomic Data and Analysis Structure (ADAS) data class {\it adf04} and will be available in OPEN-ADAS and our UK-APAP website\footnote{http://apap-network.org/}. These data can be used to improve the atomic database of astrophysical plasma codes like CHIANTI \citep{der97,der19} and SPEX \citep{kaa96,kaa18} where no data or less accurate data were available. The ordinary collision strength data will also be archived in OPEN-ADAS\footnote{http://open.adas.ac.uk/}.

\section{Discussion}
\label{sct:dis}
We selected six ions \ion{Fe}{XX}, \ion{Ca}{XIV}, \ion{Ar}{XII}, \ion{S}{X}, \ion{Si}{VIII}, and \ion{O}{II} across the iso-electronic sequence to assess the quality of our structure and collision calculations. 

\begin{figure*}
\centering
\includegraphics[width=.8\hsize, trim={1.0cm 0.5cm 1.5cm 0.5cm}, clip]{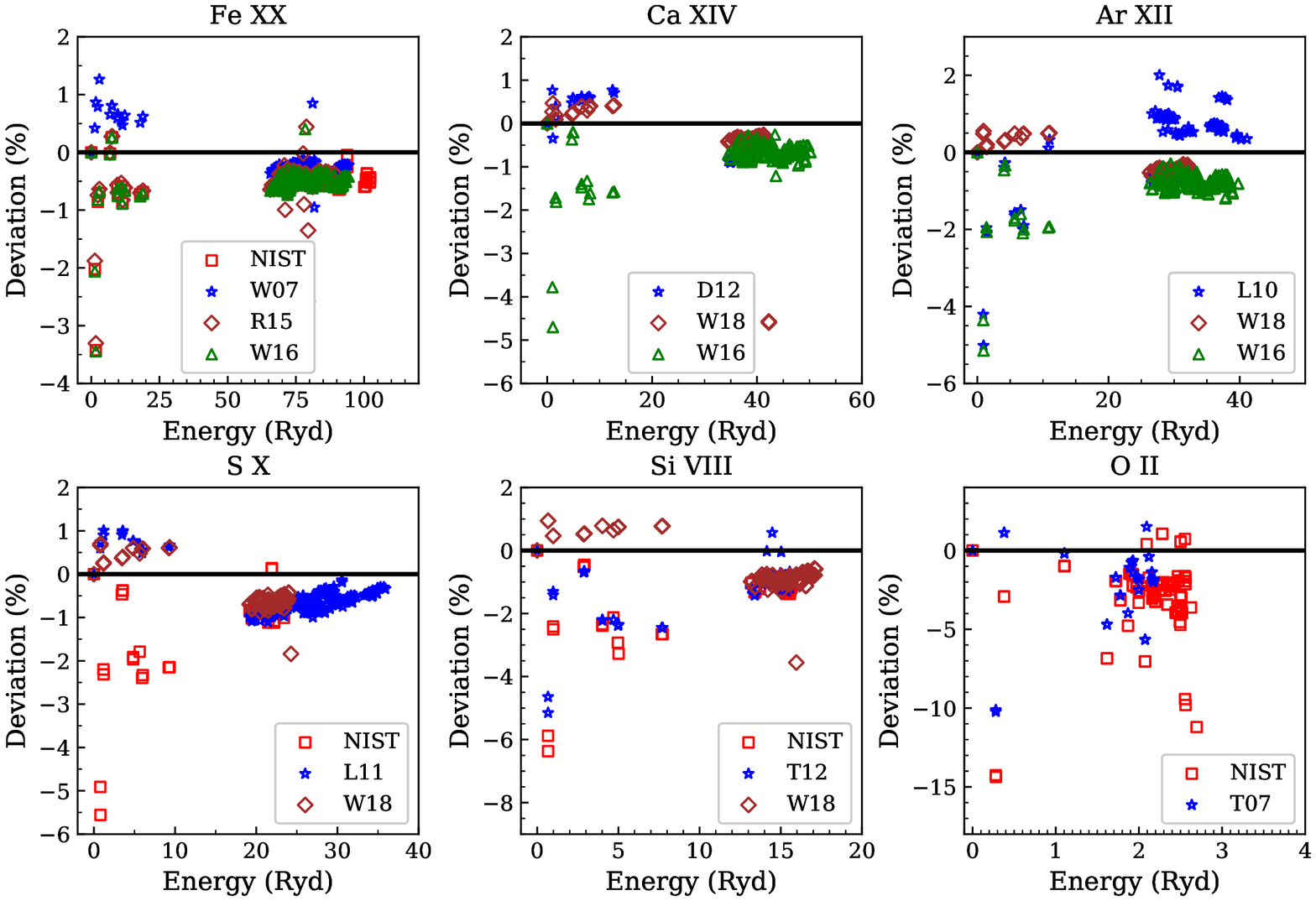}
\caption{Percentage deviations between the present energy levels (horizontal lines in black), the experimental ones (NIST) and previous works: W07 refers to \citet{wit07}, R15 refers to \citet{rad15}, W16 refers to \citet{wan16}, D12 refers to \citet{don12}, W18 refers to \citet{wan18}, L10 refers to \citet{lud10}, L11 refers to \citet{lia11}, T12 refers to \citet{tay12s}, and T07 refers to \citet{tay07}. }
\label{fig:plot_cflev}
\end{figure*}

We first compare the energy levels and transition strengths $\log (gf)$, where $g$ and $f$ are the statistical weight and oscillator strength of the transition, respectively. Fig.~\ref{fig:plot_cflev} illustrates the deviation (in percent) of the energy levels in NIST and previous works with respect to the present ones. Generally speaking, the energy levels agree to within $\sim5\%$ for the high-charge ions (e.g., \ion{Fe}{XX} and \ion{Ar}{XII}). A larger deviation ($\lesssim15\%$) is found for low-charge ions like \ion{O}{II}, in particular, for some of the low-lying energy levels.

\begin{figure*}
\centering
\includegraphics[width=.8\hsize, trim={1.0cm 0.5cm 1.5cm 0.5cm}, clip]{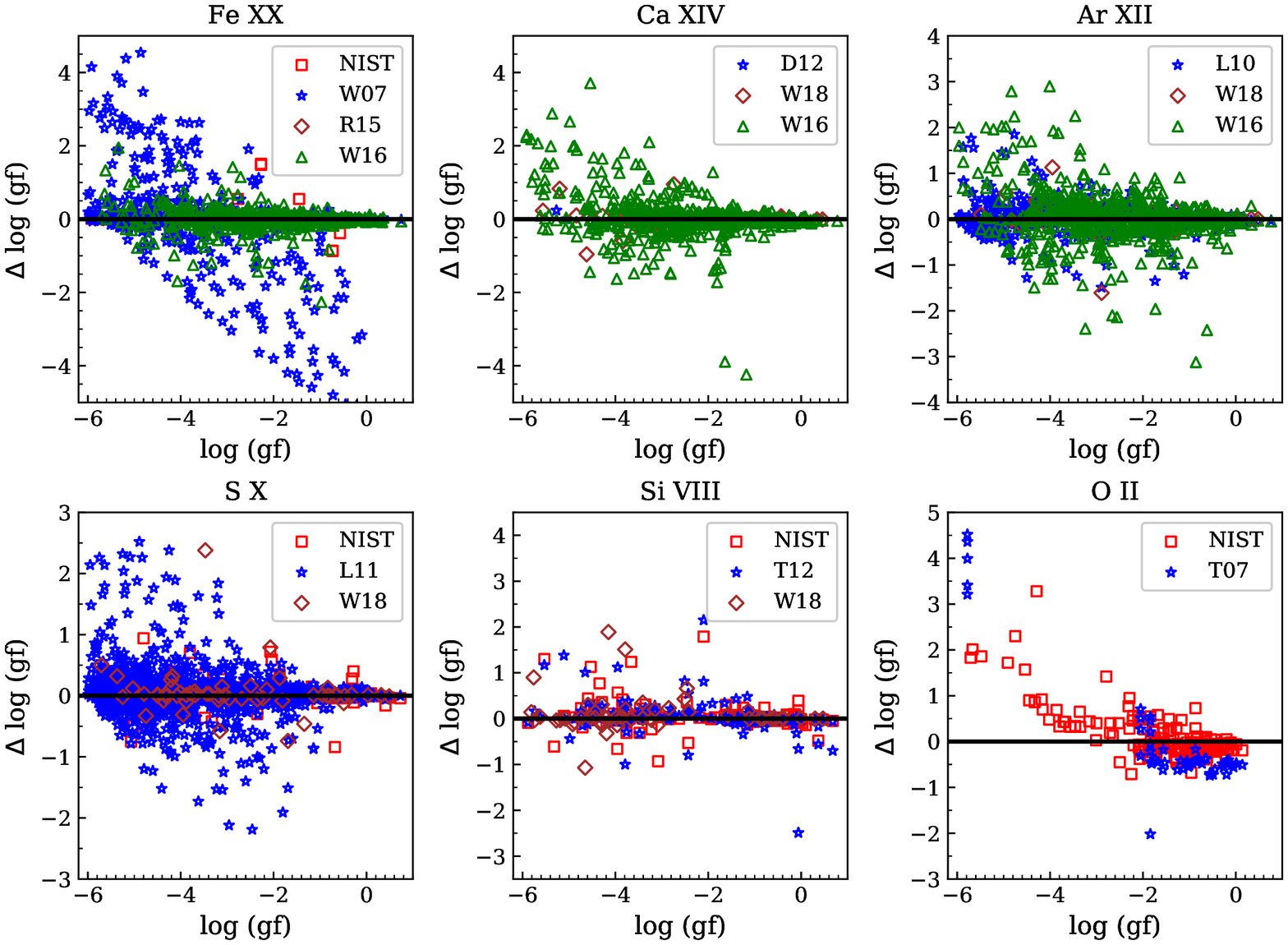}
\caption{Comparisons of $\log~(gf)$ from the present work (black horizontal line) with the experimental ones (NIST) and previous works: W07 refers to \citet{wit07}, R15 refers to \citet{rad15}, W16 refers to \citet{wan16}, D12 refers to \citet{don12}, W18 refers to \citet{wan18}, L10 refers to \citet{lud10}, L11 refers to \citet{lia11}, T12 refers to \citet{tay12s}, and T07 refers to \citet{tay07}. We note that this comparison is limited to relatively strong transitions with $\log~(gf) \gtrsim 10^{-6}$ originating from the lowest five energy levels.}
\label{fig:plot_cftran}
\end{figure*}

Fig.~\ref{fig:plot_cftran} shows the deviation of transition strengths $\Delta \log~(gf)$ in archival databases and previous works with respect to the present work. We limit the comparison to relatively strong transitions with $\log~(gf) \gtrsim 10^{-6}$ from the lowest five energy levels of the ground configuration: $2s^22p^3 ~(^4S_{3/2},^2D_{3/2,~5/2},^2P_{1/2,~3/2})$. Given the relatively low density of astrophysical plasmas, the ionic level population is dominated by the ground and first four metastable levels \citep{mao17}. Weak transitions are not expected to significantly impact the astrophysical plasma diagnostics. 

Subsequently, we compare the collision data for \ion{Fe}{XX} (Section~\ref{sct:072620}), \ion{Ca}{XIV} (Section~\ref{sct:072014}), \ion{Ar}{XII} (Section~\ref{sct:071812}), \ion{S}{XI} (Section~\ref{sct:071610}), \ion{Si}{VIII} (Section~\ref{sct:071408}), and \ion{O}{II} (Section~\ref{sct:070802}). $R$-matrix ICFT calculations were performed previously for \ion{Fe}{XX} \citep{wit07}, \ion{Ca}{XIV} \citep{wan18}, \ion{Ar}{XII} \citep{lud10, wan18}, \ion{S}{XI} \citep{lia11, wan18}, and \ion{Si}{VIII} \citep{wan18}. In addition, calculations were performed previously for \ion{Ca}{XIV} with the Dirac atomic $R$-matrix code \citep{don12}, \ion{Si}{VIII} with B-spline $R$-matrix \citep{tay12s},  and \ion{O}{II} with B-spline $R$-matrix \citep{tay07} and Breit-Pauli $R$-matrix with pseudo-states \citep{kis09}.

We use hexbin plots \citep{car87} to compare the effective collision strengths from the present work with the latest large-scale $R$-matrix calculations in the literature for \ion{Fe}{XX}, \ion{Ar}{XII}, and \ion{S}{X}. Table~\ref{tbl:cf_ecs_stat} provides some statistics of the hexbin plot comparison. Generally speaking, smaller deviations are found at higher temperatures. Since the present work has a significantly larger close-coupling expansion (725 levels), the additional resonances contribute most to the deviation at low and intermediate temperatures. Similar behaviour was noted also by \citet{fme16}. 

\begin{table*}
\caption[]{Statistics of the effective collision strength comparison for \ion{Fe}{XX}, \ion{Ar}{XII}, and \ion{S}{X}. Columns \#2--\#4 give the number of transitions with $\log (\Upsilon) > -5$ in both data sets and the percentage of transitions with deviation larger than 0.2 dex at three temperatures (ion-dependent) used for the hexbin plots. Columns \#5--\#7 are the statistics when limiting the transitions from the lowest five transitions (i.e. the ground and the first metastable levels). }
\label{tbl:cf_ecs_stat}
\centering
\begin{tabular}{c|ccc|ccc}
\hline\hline
\noalign{\smallskip} 
Ion & $T({\rm low})$ & $T({\rm middle})$ & $T({\rm high})$ & $T({\rm low})$ & $T({\rm middle})$ & $T({\rm high})$ \\
\noalign{\smallskip} 
\hline
\noalign{\smallskip} 
\ion{Fe}{XX} & $\sim41000$ ($57\%$) & $\sim38000$ ($58\%$) & $\sim35000$ ($32\%$) & $1480$ ($23\%$) & $1474$ ($27\%$) & $1398$ ($8\%$)  \\
\noalign{\smallskip} 
\ion{Ar}{XII} & $\sim12000$ ($63\%$) & $\sim15000$ ($57\%$) & $\sim15000$ ($46\%$) & $771$ ($44\%$) & $889$ ($29\%$) & $888$ ($23\%$)  \\
\noalign{\smallskip} 
\ion{S}{X} & $3438$ ($80\%$) & $3375$ ($74\%$) & $3222$ ($53\%$) & $405$ ($43\%$) & $402$ ($33\%$) & $394$ ($17\%$)  \\
\noalign{\smallskip} 
\hline
\end{tabular}
\end{table*}

When limiting the comparison to transitions from the lowest five energy levels (i.e., the ground and first four metastable levels), smaller deviations are found at all temperatures. $R$-matrix calculations without pseudo-states (including the present work) are not converged for the high-lying levels, both with respect to the $N$-electron target configuration interaction expansion and the ($N+1$)-electron close-coupling expansion. Therefore, the effective collision strengths obtained in the present and previous works involving high-lying energy levels are not converged. To improve the accuracy of transitions involving the high-lying levels with $n\ge4$, especially between these high-lying levels, larger-scale $R$-matrix ICFT calculations or $R$-matrix calculations with pseudo-state calculations are required. 

For \ion{Fe}{XX}, \ion{Ca}{XIV}, \ion{Ar}{XII}, \ion{S}{X}, \ion{Si}{VIII}, and \ion{O}{II}, we also compare selected prominent allowed and forbidden transitions (Table~\ref{tbl:gml_tbl1}) from the ground and metastable levels. Most of these transitions are used to measure the density of the solar atmosphere \citep{moh03, dza18a}. In many cases, effective collision strengths for these density diagnostic lines agree well between the present and previous works. 

\begin{table}
\caption[]{Selected prominent transitions from the lowest three energy levels for \ion{Fe}{XX}, \ion{Ca}{XIV}, \ion{Ar}{XII}, \ion{S}{X}, \ion{Si}{VIII}, and \ion{O}{II}. The rest-frame wavelengths (\AA) are taken from the CHIANTI atomic database. Forbidden transitions are labeled with (f). }
\label{tbl:gml_tbl1}
\centering
\begin{tabular}{llll}
\hline\hline
\noalign{\smallskip} 
Ion & Lower level & Upper level & $\lambda_0$ ($\AA$) \\
\noalign{\smallskip} 
\hline
\noalign{\smallskip} 
\ion{Fe}{XX} & $2s^2 2p^3~(^4S_{3/2})$ & $2s^2 2p^23d~ (^4P_{3/2})$ & 12.83 \\
& $2s^2 2p^3~(^2D_{3/2})$ & $2s^2 2p^2 3d~(^2D_{5/2})$ & 12.98 \\ 
& $2s^2 2p^3~(2D_{5/2})$ & $2s^2 2p^2 3d~(^2F_{7/2})$ & 13.09 \\
\noalign{\smallskip} 
\hline
\noalign{\smallskip} 
\ion{Ca}{XIV} & $2s^2 2p^3~(^4S_{3/2})$ & $2s 2p^4~(^4P_{5/2})$ & 193.87 \\
& $2s^2 2p^3~(^2D_{5/2})$ & $2s 2p^4~(^2D_{5/2})$ & 166.96 \\
& $2s^2 2p^3~(^2D_{5/2})$ & $2s 2p^4~(^2P_{3/2})$ & 134.27 \\
& $2s^2 2p^3~(^4S_{3/2})$ & $2s^2 2p^3~(^2D_{3/2})$ & 943.59 (f) \\
& $2s^2 2p^3~(^4S_{3/2})$ & $2s^2 2p^3~(^2D_{5/2})$ & 880.40 (f) \\
\noalign{\smallskip} 
\hline
\noalign{\smallskip} 
\ion{Ar}{XII} & $2s^2 2p^3~(^4S_{3/2})$ & $2s 2p^4~(^2P_{5/2})$ & 224.25 \\
& $2s^2 2p^3~(^2D_{5/2})$ & $2s 2p^4~(^2D_{5/2})$ & 193.70 \\
& $2s^2 2p^3~(^2D_{5/2})$ & $2s 2p^4~(^2P_{3/2})$ & 154.42 \\
& $2s^2 2p^3~(^4S_{3/2})$ & $2s^2 2p^3~(^2D_{3/2})$ & 1054.69 (f) \\
& $2s^2 2p^3~(^4S_{3/2})$ & $2s^2 2p^3~(^2D_{5/2})$ & 1018.72 (f) \\
\noalign{\smallskip} 
\hline
\noalign{\smallskip} 
\ion{S}{X} & $2s^2 2p^3~(^4S_{3/2})$ & $2s 2p^4~(^2P_{5/2})$ & 264.23 \\
& $2s^2 2p^3~(^2D_{5/2})$ & $2s 2p^4~(^2D_{5/2})$ & 228.69 \\
& $2s^2 2p^3~(^2D_{5/2})$ & $2s 2p^4~(^2P_{3/2})$ & 180.73 \\
& $2s^2 2p^3~(^4S_{3/2})$ & $2s^2 2p^3~(^2D_{3/2})$ & 1212.93 (f) \\
& $2s^2 2p^3~(^4S_{3/2})$ & $2s^2 2p^3~(^2D_{5/2})$ & 1196.22 (f) \\
\noalign{\smallskip} 
\hline
\noalign{\smallskip} 
\ion{Si}{VIII} & $2s^2 2p^3~(^4S_{3/2})$ & $2s 2p^4~(^4P_{5/2})$ & 319.84 \\
& $2s^2 2p^3~(^2D_{5/2})$ & $2s 2p^4~(^2D_{5/2})$ & 277.06 \\
& $2s^2 2p^3~(^2D_{3/2})$ & $2s 2p^4~(^2D_{3/2})$ & 276.85 \\
& $2s^2 2p^3~(^2D_{5/2})$ & $2s 2p^4~(^2P_{3/2})$ & 216.92 \\
& $2s^2 2p^3~(^4S_{3/2})$ & $2s^2 2p^3~(^2D_{3/2})$ & 1445.73 (f) \\
& $2s^2 2p^3~(^4S_{3/2})$ & $2s^2 2p^3~(^2D_{5/2})$ & 1440.51 (f) \\
\noalign{\smallskip} 
\hline
\noalign{\smallskip} 
\ion{O}{II} & $2s^2 2p^3~(^4S_{3/2})$ & $2s^2 2p^3~(^2D_{5/2})$ & 3729.88 (f) \\
& $2s^2 2p^3~(^4S_{3/2})$ & $2s^2 2p^3~(^2D_{3/2})$ & 3727.09 (f) \\ 
& $2s^2 2p^3~(^4S_{3/2})$ & $2s^2 2p^3~(^2P_{1/2})$ & 2470.97 (f) \\ 
& $2s^2 2p^3~(^4S_{3/2})$ & $2s^2 2p^3~(^2P_{3/2})$ & 2471.09 (f) \\
& $2s^2 2p^3~(^2D_{5/2})$ & $2s^2 2p^3~(^2P_{1/2})$ & 7320.94 (f) \\ 
& $2s^2 2p^3~(^2D_{5/2})$ & $2s^2 2p^3~(^2P_{3/2})$ & 7322.01 (f) \\ 
& $2s^2 2p^3~(^2D_{3/2})$ & $2s^2 2p^3~(^2P_{1/2})$ & 7331.69 (f) \\ 
& $2s^2 2p^3~(^2D_{3/2})$ & $2s^2 2p^3~(^2P_{3/2})$ & 7332.76 (f) \\ 
\noalign{\smallskip} 
\hline
\end{tabular}
\end{table}

\subsection{\ion{Fe}{XX}}
\label{sct:072620}
The most recent calculation of $R$-matrix electron-impact excitation data for \ion{Fe}{XX} (or ${\rm Fe^{19+}}$) is presented by \citet[][W07 hereafter]{wit07}. We limit our comparison to W07 and refer readers to W07 for their comparison with other earlier calculations \citep{but01,mla01}. 

Both W07 and the present work use the AUTOSTRUCTURE code for the structure calculation. As shown in the top-left panel of Fig.~\ref{fig:plot_cflev}, the energy levels of the present work and W07 agree within $\lesssim1\%$. The first few levels of the present work and W07 differ up to $\sim4\%$ with respect to NIST, \citet[][R15]{rad15}, and \citet[][W16]{wan16}. The latter two more accurate structure calculations were performed with the multi-configuration Dirac-Fock theory and many-body perturbation theory, respectively. As shown in the top-left panel of Fig.~\ref{fig:plot_cftran}, the transition strengths agree well between NIST, R15, W16, and the present work with merely a few exceptions. Larger deviations are found between W07 and other works.

\begin{figure*}
\centering
\includegraphics[width=.8\hsize, trim={0.5cm 0.5cm 1.5cm 0.5cm}, clip]{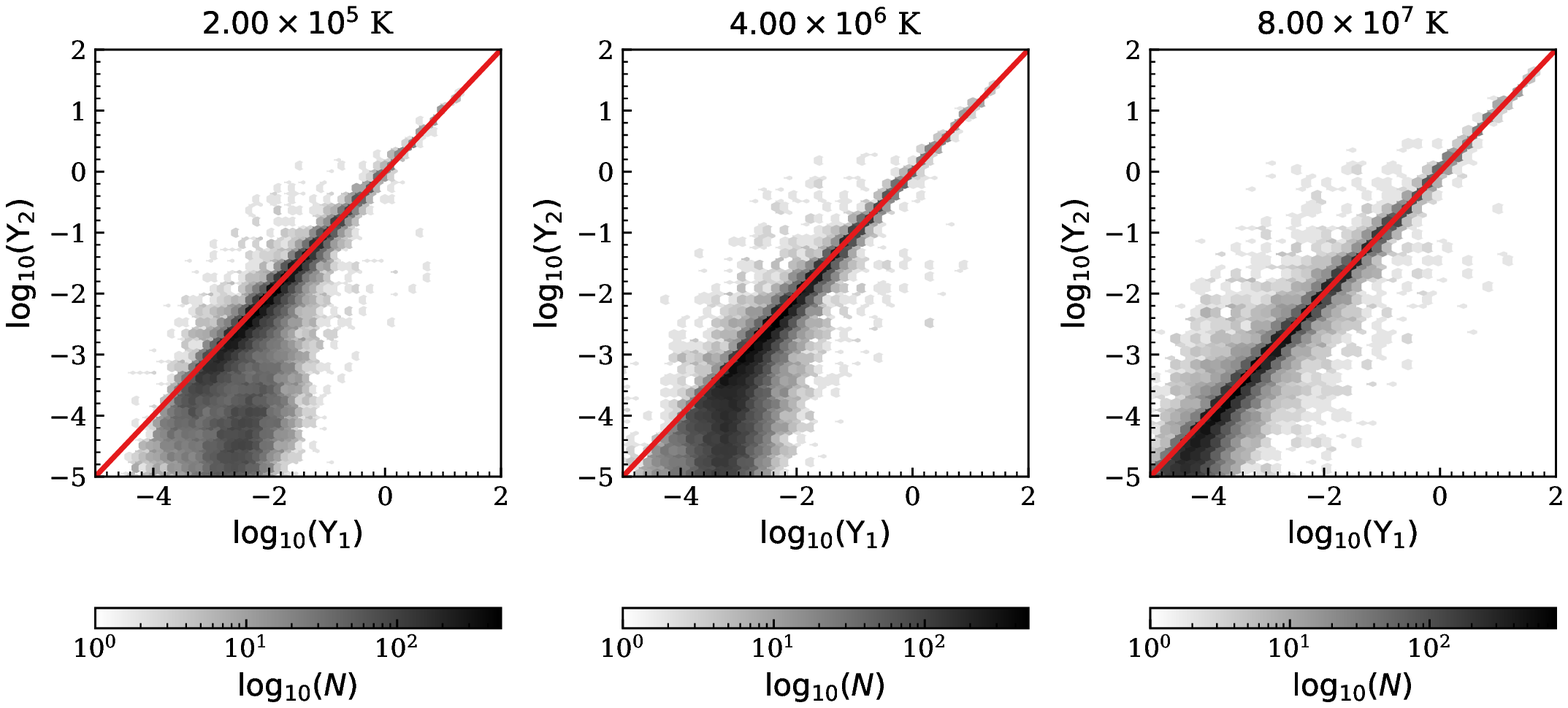}
\caption{Hexbin plots of the comparison of the \ion{Fe}{XX} (or ${\rm Fe^{20+}}$) effective collision strengths between the present work ($\Upsilon_1$) and \citet[][$\Upsilon_2$]{wit07} at $T\sim2.00\times10^5~{\rm K}$ (left) and $4.00\times10^6~{\rm K}$ (middle), and $\sim8.00\times10^7~{\rm K}$ (right). The darker the color, the greater the number of transitions $\log_{10}(N)$. The diagonal line in red indicates $\Upsilon_1=\Upsilon_2$.}
\label{fig:hb_ecs_072620}
\end{figure*}

\begin{figure}
\centering
\includegraphics[width=.8\hsize, trim={0.cm 0.5cm 1.0cm 0.5cm}, clip]{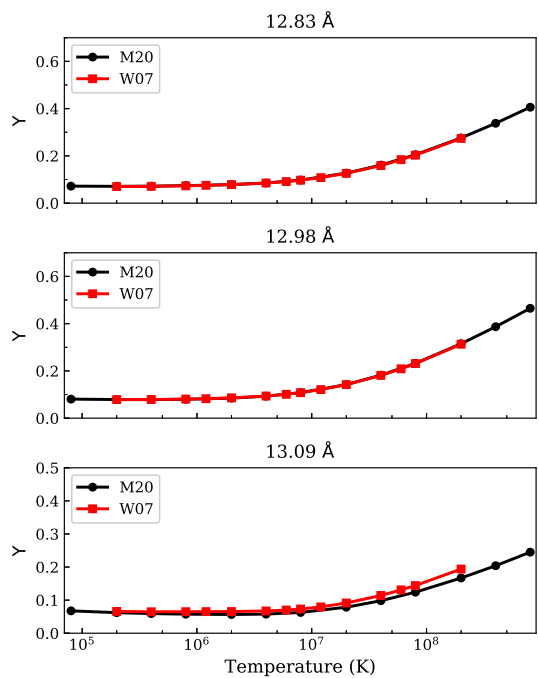}
\caption{Comparison of \ion{Fe}{XX} (or ${\rm Fe^{19+}}$) effective collision strengths between the present work (M20) and \citet[][W07]{wit07} for selected dipole transitions from the ground (upper) and metastable (middle and bottom) levels listed in Table~\ref{tbl:gml_tbl1}. }
\label{fig:plot_ecs_072620}
\end{figure}

Both W07 and the present work use the $R$-matrix ICFT method for the scattering calculation. W07 included 302 fine-structure levels in the close-coupling expansions. The present work has 725 levels. Fig.~\ref{fig:hb_ecs_072620} shows the hexbin plot comparison of the effective collision strengths at $T\sim2.00\times10^5~{\rm K}$ (left), $4.00\times10^6~{\rm K}$ (middle), and $\sim8.00\times10^7~{\rm K}$ (right). As shown in Fig.~\ref{fig:plot_ecs_072620}, the effective collision strengths for the three selected dipole transitions from the ground (12.83~\AA) and metastable (12.98~\AA\ and 13.09~\AA) levels agree well between the present work and W07.

\subsection{\ion{Ca}{XIV}}
\label{sct:072014}
The most recent $R$-matrix calculations of the electron-impact excitation data of \ion{Ca}{XIV} (or ${\rm Ca^{13+}}$) are presented in \citet[][W18]{wan18} and \citet[][D12 hereafter]{don12}. 

The general-purpose relativistic atomic structure package (GRASP) and AUTOSTRUCTURE were used by D12 and W18, respectively, for their atomic structure calculations. The energy levels and transition strengths of D12, W18, and the present work agree well with each other (the upper-middle panels of Fig.~\ref{fig:plot_cflev} and \ref{fig:plot_cftran}). 

D12 included 272 fine-structure levels for the target ion. The Dirac atomic $R$-matrix code (DARC) was used for the collision calculation. Effective collision strengths from the ground level to the lowest 15 levels are tabulated in their Table 4 and archived as supplementary data. The CHIANTI atomic database includes a few more effective collision strengths from the metastable levels provided by \citet{don12}. W18 also included 272 fine-structure levels for the target ion. Their scattering calculation was performed via the $R$-matrix ICFT method. Effective collision strengths from the ground level to the lowest 120 levels are tabulated in their Table 22 for \ion{Ca}{XIV}. 

\begin{figure}
\centering
\includegraphics[width=.87\hsize, trim={0.cm 0.5cm 0.5cm 0.5cm}, clip]{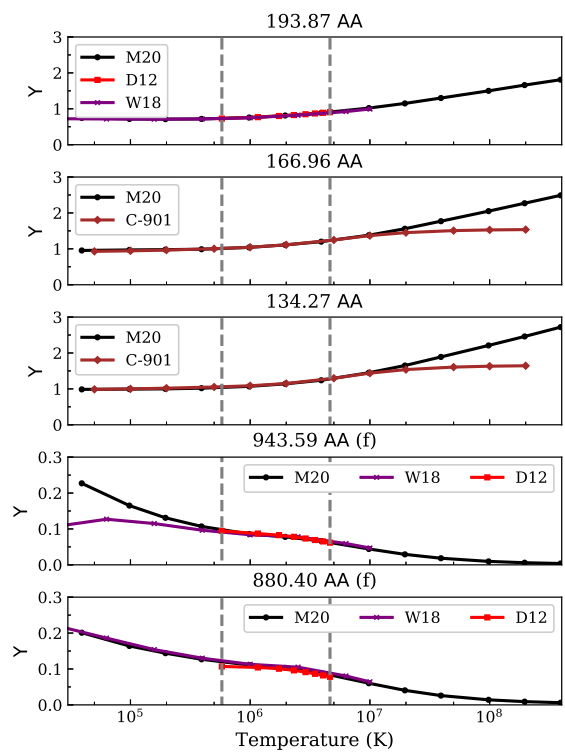}
\caption{Comparison of \ion{Ca}{XIV} (or ${\rm Ca^{13+}}$) effective collision strengths between the present work (M20), \citet[][D12]{don12}, and \citet[][W18]{wan18} for selected transitions listed in Table~\ref{tbl:gml_tbl1}. The top panel is a dipole transition from the ground level. For the two dipole metastable transitions (in the second and third panels from the top), effective collision strengths C-901 are obtained directly from the CHIANTI atomic database. The vertical dashed lines indicate the temperature range originally provided by \citet{don12}. The brown diamonds outside this temperature range are extrapolated data in CHIANTI. The bottom two panels are forbidden transitions from the ground level to the first two metastable levels. }
\label{fig:plot_ecs_072014}
\end{figure}

As shown in Fig.~\ref{fig:plot_ecs_072014}, the effective collision strengths of three selected dipole transitions from the ground and metastable levels agree well between the three data sets (D12, W18, and the present work) within the common temperature range. For the two metastable transitions (166.96~\AA\ and 134.27~\AA), the extrapolation at higher temperatures in the current version of CHIANTI atomic database (v9.0.1) was not carried out self-consistently, hence the deviations. For the forbidden transition (943.59~\AA), the effective collision strengths at $T\lesssim10^{6}$~K differ by a factor of two between the present work and W18. Good agreement is found for the other forbidden transition 880.40~\AA\ between the present work, D12, and W18.

\begin{figure}
\centering
\includegraphics[width=.7\hsize,  angle=90, clip]{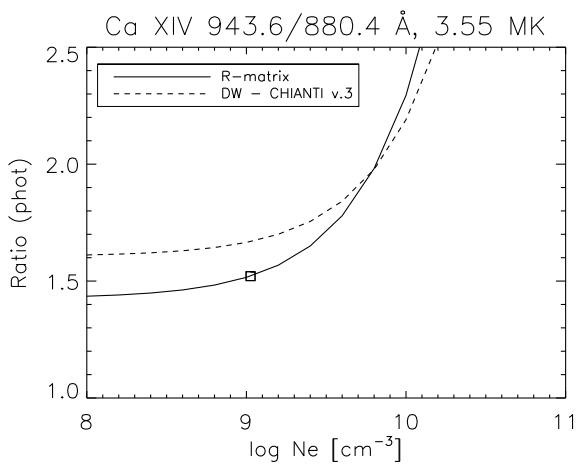}
\caption{Line ratios (in ${\rm phot~cm^{-2}~s^{-1}}$) for key diagnostics lines of \ion{Ca}{XIV} as a function of density at a fixed temperature of $3.55\times10^6$~K (the peak ion abundance in ionization equilibrium). 
The solid curve is calculated with the present $R$-matrix data, while the dashed curve use  distorted wave data as incorporated in CHIANTI version 3. The square indicates the measurement from a 
post-flare solar SUMER observation \citep{lan03}.  }
\label{fig:ca_14_ratio}
\end{figure}

We note that the two forbidden lines from the $^2D_{5/2,3/2}$ to the ground state for Ca {\sc xiv}, observed at 880.4~\AA\ and 943.6~\AA\ respectively, are useful density diagnostics, being relatively close in wavelength. These lines have been observed with several solar instruments, most notably the SUMER  (Solar Ultraviolet Measurements of Emitted Radiation) spectrograph on SOHO (Solar and Heliospheric Observatory, see e.g. \cite{cur04}.  \cite{lan03} reported SUMER observations of post-flare loops and noted significant discrepancies (factors of up to 10) between the densities obtained from different ions. They used CHIANTI version 3 atomic data, which included DW rates for Ca {\sc xiv} and \ion{Ar}{XII}. For Ca {\sc xiv}, no density was obtained, as the observed ratio was below the low-density limit, as shown in Fig.~\ref{fig:ca_14_ratio}. To provide an application of the present atomic rates in solar observations, we built a development version of CHIANTI with the present data. We obtain a density of $1.05\times10^{9}~{\rm cm^{-3}}$. 

\subsection{\ion{Ar}{XII}}
\label{sct:071812}
The most recent $R$-matrix calculations of electron-impact excitation data for \ion{Ar}{XII} (or ${\rm Ar^{11+}}$) are presented in \citet[][W18]{wan18} and \citet[][L10 hereafter]{lud10}. 

L10, W18 and the present work all used AUTOSTRUCTURE for the atomic structure calculation. As shown in the upper-right panel of Fig.~\ref{fig:plot_cflev}, the energies of the low-lying levels in L10 agree well with \citet[][W16]{wan16}, which was calculated with the many-body perturbation theory. The energies of the high-lying levels in L10 are $\sim2-3~\%$ offset with respect to W16. The level energies of W18 and the present work agree with each other to within $\sim1~\%$, with up to $\sim5~\%$ deviation with respect to W16 for the low-lying transitions. The transition strengths of L10, W16, W18, and the present work agree well with each other (the upper-right panel of Fig.~\ref{fig:plot_cftran}). 

\begin{figure*}
\centering
\includegraphics[width=.8\hsize, trim={0.5cm 0.5cm 1.5cm 0.5cm}, clip]{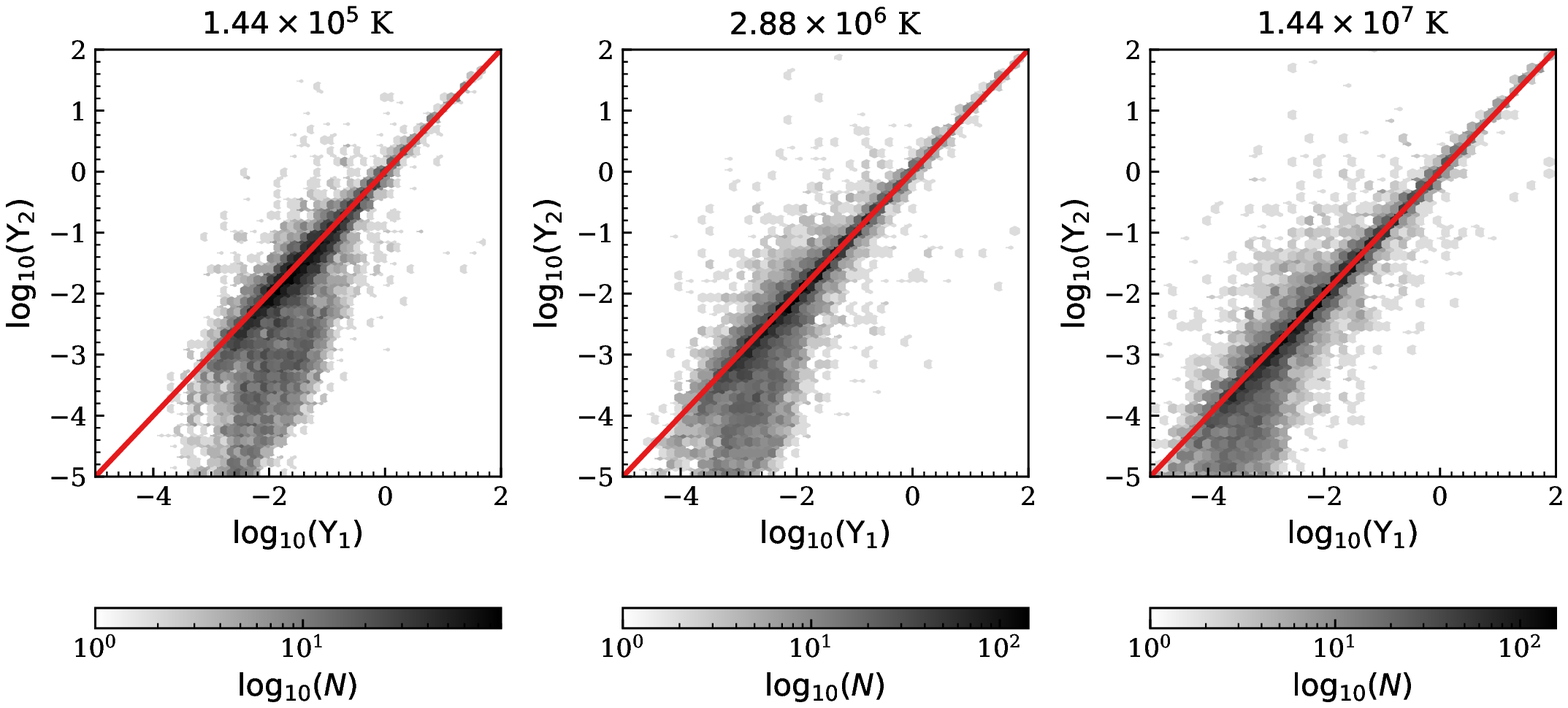}
\caption{Hexbin plots of the comparison of the \ion{Ar}{XII} (or ${\rm Ar^{11+}}$) effective collision strengths between the present work ($\Upsilon_1$) and \citet[][$\Upsilon_2$]{lud10} at $T\sim1.44\times10^5~{\rm K}$ (left) and $2.88\times10^6~{\rm K}$ (middle), and $\sim1.44\times10^7~{\rm K}$ (right). The darker the color, the greater the number of transitions $\log_{10}(N)$. The diagonal line in red indicates $\Upsilon_1=\Upsilon_2$.}
\label{fig:hb_ecs_071812}
\end{figure*}

L10, W18 and the present work all used the $R$-matrix ICFT method for the scattering calculation. L10 included 186 fine-structure levels of the target ion. At the low temperature end ($2.88\times10^4-2.88\times10^5$~K), effective collision strengths for transitions involving levels \#158 to \#186 might have some issues in their post-processing of the ordinary collision strengths (see Appendix~\ref{sct:071812app}). W18 included 272 fine-structure levels of the target ion.
Effective collision strengths from the ground level to the lowest 120 levels are tabulated in their Table 28 for \ion{Ar}{XII}. Fig.~\ref{fig:hb_ecs_071812} shows the hexbin plot comparison of the effective collision strengths at $T\sim1.44\times10^5~{\rm K}$ (left) and $2.88\times10^6~{\rm K}$ (middle), and $\sim1.44\times10^7~{\rm K}$ (right). 

\begin{figure}
\centering
\includegraphics[width=.87\hsize, trim={0.cm 0.5cm 0.5cm 1.2cm}, clip]{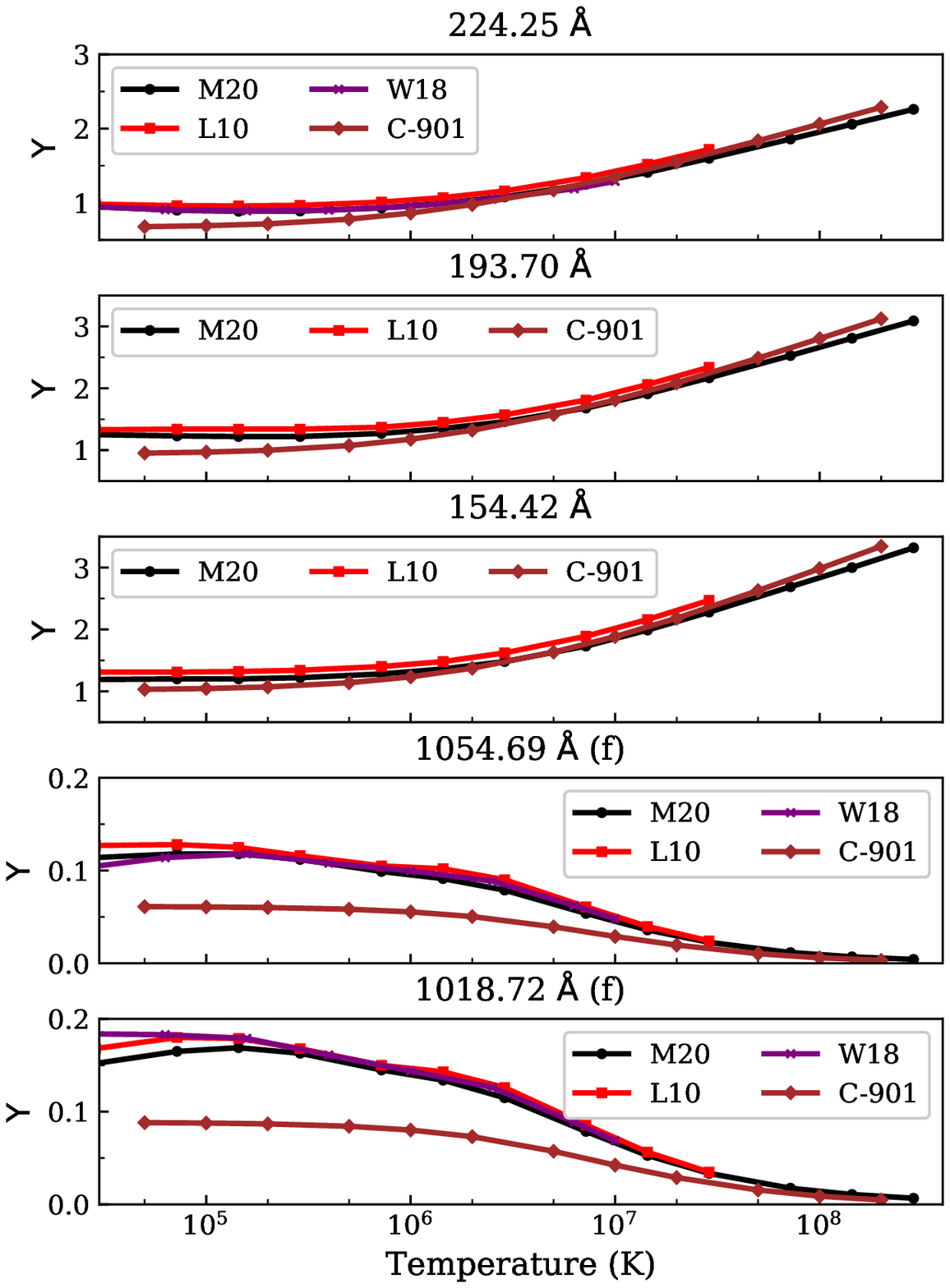}
\caption{Comparison of \ion{Ar}{XII} (or ${\rm Ar^{11+}}$) effective collision strengths between the present work (M20), \citet[][L10]{lud10}, \citet[][W18]{wan18} and \citet[][distorted wave]{eis05} as incorporated in the CHIANTI atomic database v9.0.1 (C-901) for selected transitions listed in Table~\ref{tbl:gml_tbl1}. The top panel is a dipole transition from the ground level, followed by two metastable transitions. The bottom two panels are forbidden transitions from the ground level to the first two metastable levels. }
\label{fig:plot_ecs_071812}
\end{figure}

As shown in Fig.~\ref{fig:plot_ecs_071812}, the effective collision strengths of three selected dipole transitions from the ground and metastable levels agree well between the four data sets: present work (M20), \citet[][L10]{lud10}, \citet[][W18]{wan18} and \citet[][distorted wave]{eis05} as incorporated in the CHIANTI atomic database v9.0.1 (C-901). For the forbidden transitions from the ground to the first two metastable levels (1054.69~\AA\ and 1018.72~\AA), while the three $R$-matrix data sets agree better with each other, the distorted wave (DW) data set \citep{eis05} as incorporated in the CHIANTI atomic database v9.0.1 differs by a factor of two at $T\lesssim10^7$~K. 

As for Ca {\sc xiv}, we built a development version of CHIANTI with the present data of Ar {\sc xii}. The radiative data in the public version v9.0.1 originated from \citet{eis05}. In the development version, we use the A-values (i.e. transition probabilities) of the present work with the exception of the transition $2s^22p^3 (^2D_{5/2})$ to $2s^22p^3 (^2D_{3/2})$, as the A-value of this transition in the present work is 0.77, a factor of $\sim2$ larger than that of a multi-configuration Dirac-Fock calculation from C. Froese Fischer\footnote{https://nlte.nist.gov/MCHF/Elements/Ar/N\_18.32.mcdhfSD-lin.dat.mp} and \citet{eis05}. The rest of the A-values agree with these two sources to within 20~\%. 

\begin{figure}
\centering
\includegraphics[width=.8\hsize, trim={1.0cm 0.5cm 0.5cm 0.cm}, clip]{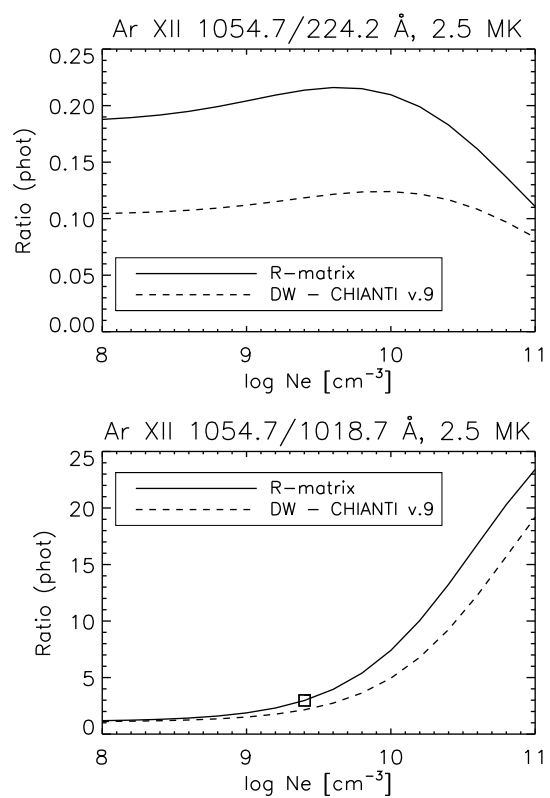}
\caption{Line ratios (in ${\rm phot~cm^{-2}~s^{-1}}$) for key diagnostic lines of \ion{Ar}{XII} as a function of density at a fixed temperature of $2.5\times10^6$~K. The solid curve is calculated with the present $R$-matrix data, while the dashed curve used the distorted wave data of \cite{eis05}, as incorporated in CHIANTI v9.0.1. The upper panel is the line ratio between the forbidden transition at 1054.7~\AA\ and the resonance transition at 224.2~\AA, while the lower panel is the line ratio between two forbidden transitions (Table~\ref{tbl:gml_tbl1}). The measurement (square) in the lower panel is from a post-flare solar SUMER observation \citep{lan03}. }
\label{fig:ar_12_ratios}
\end{figure}

Within the $2s^22p^3$ ground configuration, the two forbidden transitions at 1018.72~\AA\ and 1054.69~\AA\ (Table~\ref{tbl:gml_tbl1}) are the most important plasma diagnostics lines. These two UV lines have been observed by several solar instruments, most notably with SOHO/SUMER). They are potentially very useful to measure the solar Ar abundance. Due to the lack of photospheric lines, the solar Ar abundance cannot be measured directly \citep{lod08}. It can be derived indirectly from solar wind measurements by comparing line intensities of Ar with those from other elements. According to Fig.~\ref{fig:plot_ecs_071812}, we expect large difference in the line ratios of the development and public versions of CHIANTI. The top panel of Fig.~\ref{fig:ar_12_ratios} shows the line ratio between the forbidden transition at 1054.69~\AA\ (the stronger of the two) and the resonance transition at 224.2~\AA\ (Table~\ref{tbl:gml_tbl1}). As the resonance transition is mainly populated by direct excitation from the ground level via a strong dipole allowed transition, large differences between the distorted wave and the $R$-matrix ratios are not seen (Fig.~\ref{fig:plot_ecs_071812}). On the other hand, the increase in the effective collision strength (Fig.~\ref{fig:plot_ecs_071812}) leads to the increase of nearly a factor of two in the line ratio. 

The two forbidden lines are also very useful to measure electron densities in active regions, as they are close in wavelength. The lower panel of Fig.~\ref{fig:ar_12_ratios} shows the line ratio between the two forbidden transitions as a function of density at a fixed temperature of $2.5\times10^6$~K (the peak of ion abundance in ionization equilibrium). It is clear that significant differences in the theoretical ratio are present at higher densities, typical of active regions and flares. We also show the measurement  by \citet{lan03} from SUMER observations of active region post-flare loops. The two lines were observed within 5 minutes. The derived density we obtain is $2.9\times10^{9}~{\rm cm^{-3}}$, nearly a factor of two lower than that obtained with the DW data, and in good agreement with the density we have obtained from the \ion{Ca}{XIV} lines, considering the different formation temperature of the two ions. 

Two other strong forbidden transitions within the ground configuration have also been observed in the EUV by SUMER \citep{cur04}. They are the decays to the ground state from the $^2P_{3/2,1/2}$ levels, at 649.1~\AA\ and 670.3~\AA, respectively. They are in 
principle also useful density diagnostics, as they are close in wavelength. However, the  649.1~\AA\ line is blended with a Si {\sc x} transition. These lines were observed one hour apart from the 1018.7, 1054.7~\AA\ lines \citep{lan03}, hence their intensities cannot be directly compared.

\subsection{\ion{S}{X}}
\label{sct:071610}
The most recent $R$-matrix calculations of electron-impact excitation data for \ion{S}{X} (or ${\rm S^{9+}}$) are presented in \citet[][W18]{wan18} and \citet[][L11 hereafter]{lia11}. 

L11, W18 and the present work all used AUTOSTRUCTURE for the atomic structure calculation. As shown in the bottom-left panel of Fig.~\ref{fig:plot_cflev}, the level energies of L11, W18, and the present work agree with each other within $\sim1-2~\%$ with up to $\sim6-7~\%$ deviation with respect to NIST for the low-lying transitions. The transition strengths of NIST, L11, W18, and the present work agree well with each other (the bottom-left panel of Fig.~\ref{fig:plot_cftran}). 

\begin{figure*}
\centering
\includegraphics[width=.8\hsize, trim={0.5cm 0.5cm 1.5cm 0.5cm}, clip]{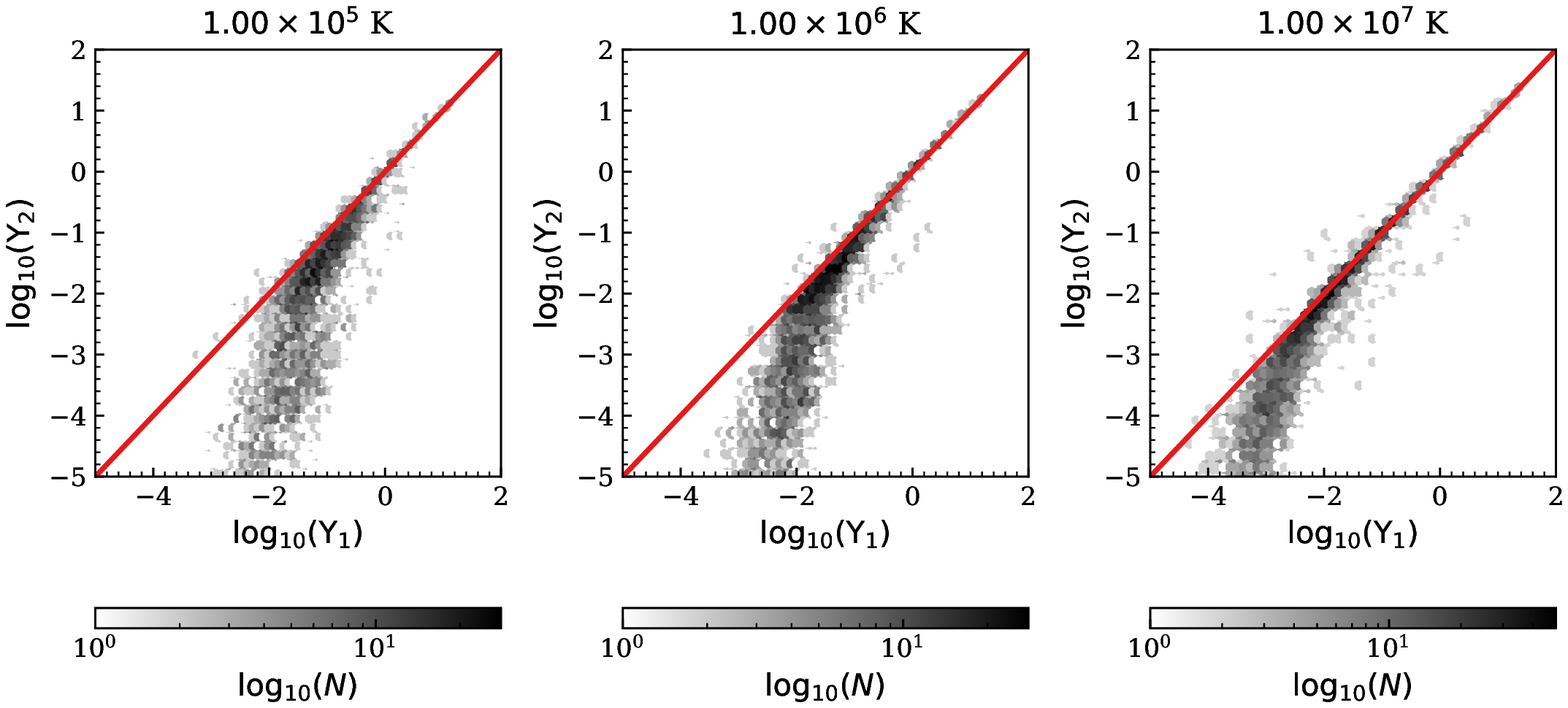}
\caption{Hexbin plots of the comparison of the \ion{S}{X} (or ${\rm S^{9+}}$) effective collision strengths between the present work ($\Upsilon_1$) and \citet[][$\Upsilon_2$]{lia11} at $T\sim1.00\times10^5~{\rm K}$ (left) and $1.00\times10^6~{\rm K}$ (middle), and $\sim1.00\times10^7~{\rm K}$ (right). The darker the color, the greater the number of transitions $\log_{10}(N)$. The diagonal line in red indicates $\Upsilon_1=\Upsilon_2$.}
\label{fig:hb_ecs_071610}
\end{figure*}

L11, W18 and the present work all used the $R$-matrix ICFT method for the scattering calculation. L11 included 84 fine-structure levels of their effective collision strengths. W18 included 272 fine-structure levels of the target ion. Effective collision strengths from the ground level to the lowest 120 levels are tabulated in their Table 26 for \ion{S}{X}. Fig.~\ref{fig:hb_ecs_071610} shows the hexbin plot comparison of the effective collision strengths at $T\sim1.00\times10^5~{\rm K}$ (left) and $1.00\times10^6~{\rm K}$ (middle), and $\sim1.00\times10^7~{\rm K}$ (right). 

\begin{figure}
\centering
\includegraphics[width=.87\hsize, trim={0.cm 0.5cm 0.5cm 0.5cm}, clip]{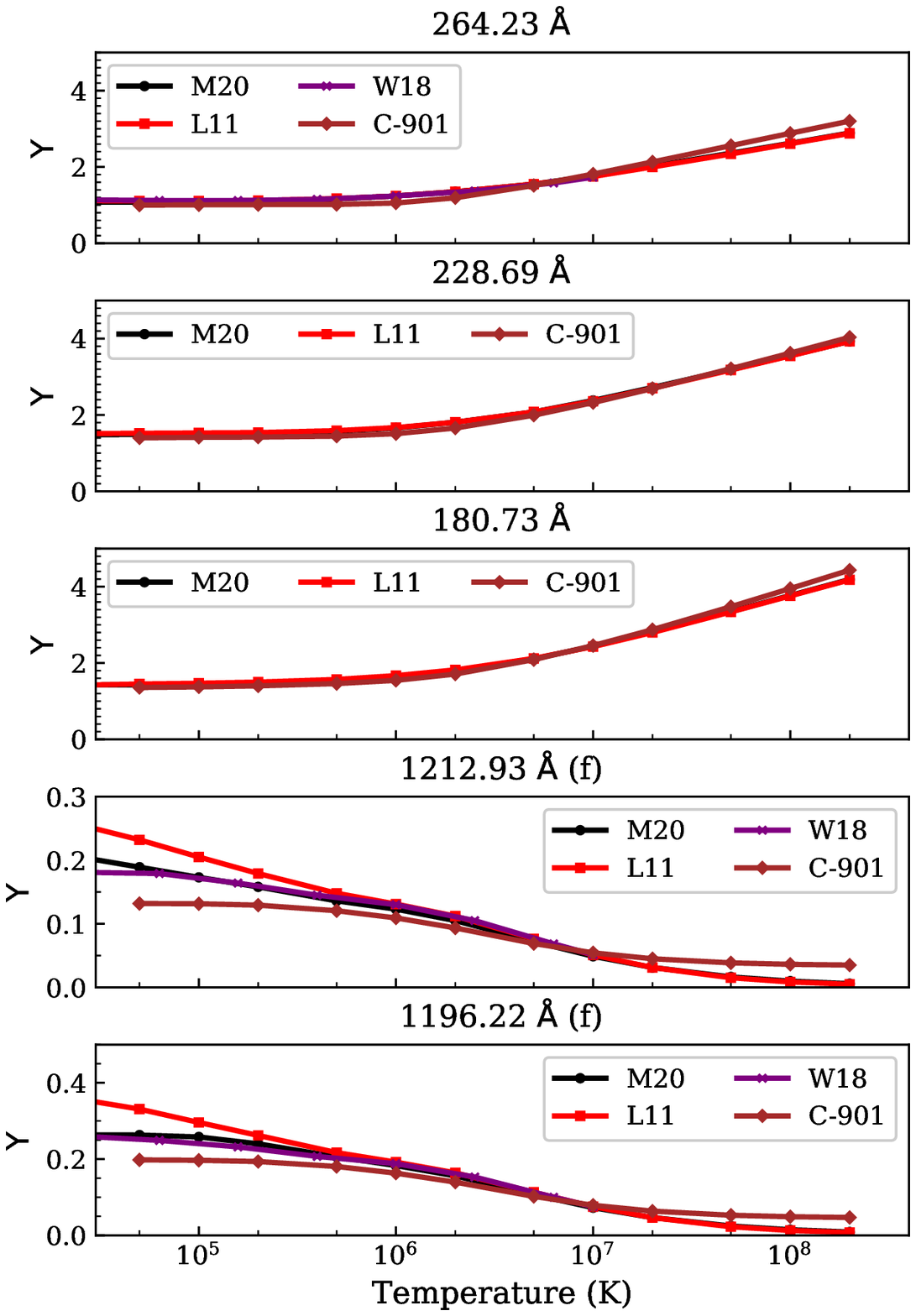}
\caption{Comparison of \ion{S}{X} (or ${\rm S^{9+}}$) effective collision strengths between the present work (M20), \citet[][L11]{lia11}, \citet[][W18]{wan18} and \citet[][$R$-matrix]{bel00} as incorporated in the CHIANTI atomic database v9.0.1 (C-901) for selected transitions listed in Table~\ref{tbl:gml_tbl1}. The top panel is a dipole transition from the ground level, followed by two metastable transitions. The bottom two panels are forbidden transitions from the ground level to the first two metastable levels. }
\label{fig:plot_ecs_071610}
\end{figure}

As shown in Fig.~\ref{fig:plot_ecs_071610}, the effective collision strengths of three selected dipole transitions from the ground and metastable levels agree well between the four data sets: present work (M20), \citet[][L11]{lia11}, \citet[][W18]{wan18}, and \citet[][$R$-matrix]{bel00} as incorporated in the CHIANTI atomic database v9.0.1 (C-901). For the two forbidden transitions from the ground to the first two metastable levels (1212.93~\AA\ and 1196.22~\AA), at $T\lesssim10^6$~K, the present work agrees with W18, while L11 is larger by a factor of two and C-901 is smaller by a factor of two. At $T\gtrsim10^7$~K, the present work agrees with L11, while C-901 is larger by an order of magnitude.

\subsection{\ion{Si}{VIII}}
\label{sct:071408}
The most recent $R$-matrix calculations of  electron-impact excitation data for \ion{Si}{VIII} (or ${\rm Si^{7+}}$) are presented in \citet[][W18]{wan18} and \citet[][T12]{tay12s}. 

Both W18 and the present work used AUTOSTRUCTURE for the atomic structure calculation, while T12 used the multi-configuration Hartree-Fock method. As shown in the bottom-middle panel of Fig.~\ref{fig:plot_cflev}, the level energies of W18 and the present work agree with each other within $\sim1-2~\%$ with up to $\sim8~\%$ deviation with respect to NIST and T12 for the low-lying transitions. The transition strengths of NIST, T12, W18, and the present work agree well with each other (the bottom-middle panel of Fig.~\ref{fig:plot_cftran}). 

Both W18 and the present work used the $R$-matrix ICFT method for the scattering calculation, while T12 used the B-spline $R$-matrix method. T12 included 68 fine-structure target levels for their effective collision strengths. W18 included 272 fine-structure levels of the target ion. Effective collision strengths from the ground level to the lowest 120 levels are tabulated in their Table 26 for \ion{Si}{viii}.

\begin{figure}
\centering
\includegraphics[width=.87\hsize, trim={0.cm 0.5cm 0.5cm 0.5cm}, clip]{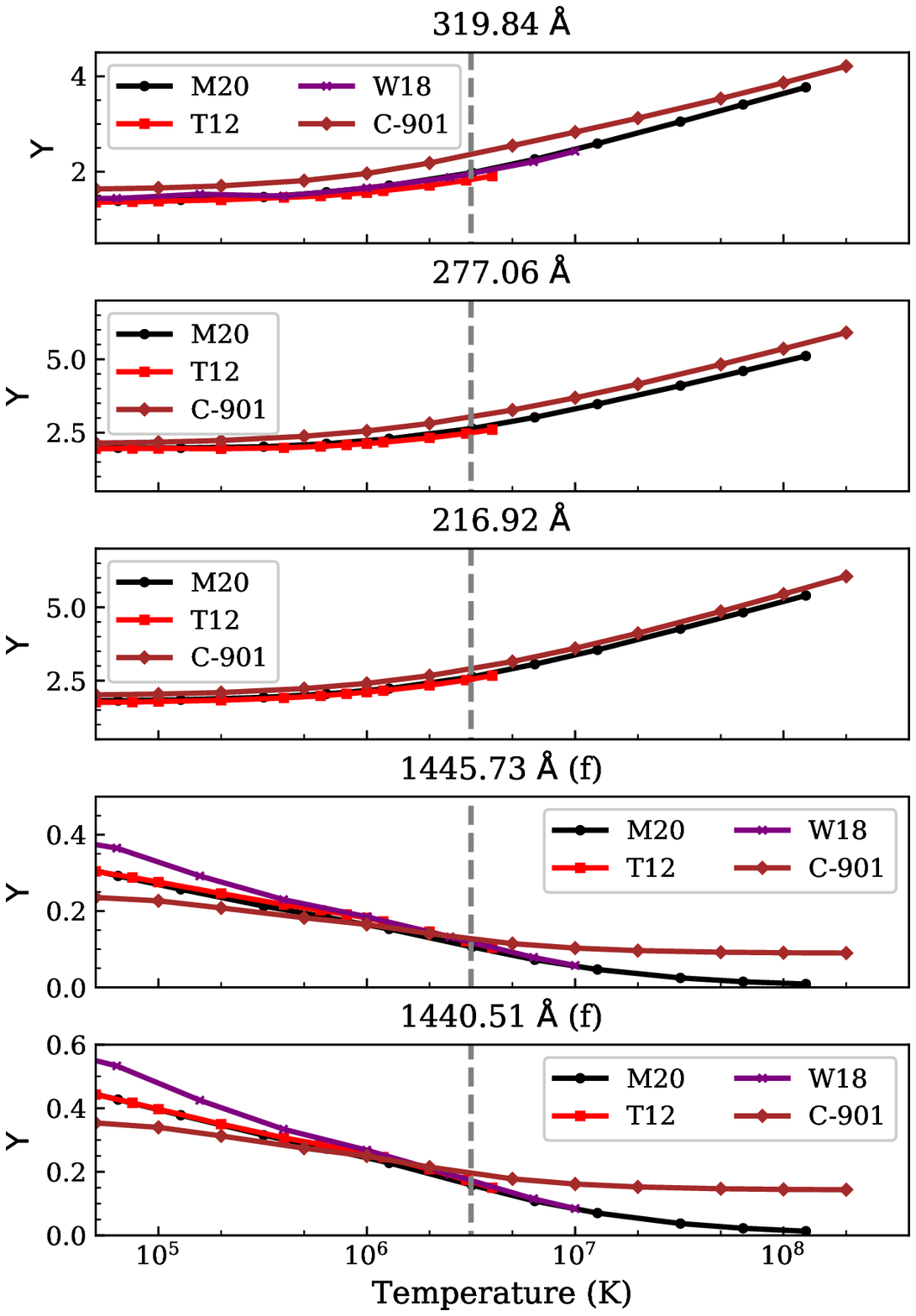}
\caption{Comparison of \ion{Si}{VIII} (or ${\rm Si^{7+}}$) effective collision strengths between the present work (M20), \citet[][T12]{tay12s}, \citet[][W18]{wan18}, and \citet{bel01} as incorporated in the CHIANTI atomic database v9.0.1 (C-901) for selected transitions listed in Table~\ref{tbl:gml_tbl1}. The top panel is a dipole transition from the ground level, followed by two metastable transitions. The bottom two panels are forbidden transitions from the ground level to the first two metastable levels. The vertical dashed lines indicate the upper temperature limit originally provided by \citet{bel01}. The brown diamonds beyond this temperature limit are extrapolated in CHIANTI. }
\label{fig:plot_ecs_071408}
\end{figure}

In Fig.~\ref{fig:plot_ecs_071408}, we compare the effective collision strengths of selected transitions listed in Table~\ref{tbl:gml_tbl1}. The values for the three dipole transitions from the ground and metastable levels agree well between the three data sets: present work (M20), \citet[][T12]{tay12s}, \citet[][W18]{wan18}. The $R$-matrix data set of \citet{bel01} as incorporated in the CHIANTI atomic database v9.0.1 (C-901) is also comparable to the other $R$-matrix data sets. For the two forbidden transitions from the ground to the first two metastable levels (1445.73~\AA\ and 1440.51~\AA), the present work and T12 agree better with each other. The previous $R$-matrix results of \citet{bel01} and \citet{wan18} are smaller and larger at $T\lesssim10^{6.5}$~K, respectively. Originally, the data of \citet{bel01} are provided in the temperature range of $10^{3.3-6.5}$~K. The extrapolated data of \citet{bel01} in the CHIANTI atomic database v9.0.1 is larger than the other $R$-matrix data at $T\lesssim10^{6.5}$~K. 

As in the case of \ion{Ar}{XII}, we built a development version of CHIANTI with the present data of \ion{Si}{VIII}. In the development version, we use the A-values of the present work with the exception of transitions between the $2s^22p^3$ and $2s2p^4$ configurations, where values from a multi-configuration Hartree-Fock calculation by \citet{tac02} were used. In the public version of CHIANTI (v9.0.1), the A-values draws from several sources \citep{mer99,zha99,bha03a}. The effective collision strengths use the $R$-matrix data of \citet{bel01} for the ground configuration and distorted wave data \citep{zha99,bha03a} for the rest. 

\begin{figure}[!htbp]
\centering
\includegraphics[width=.75\hsize, trim={0.cm 0.5cm 0.5cm 0.5cm},clip, angle=-90]{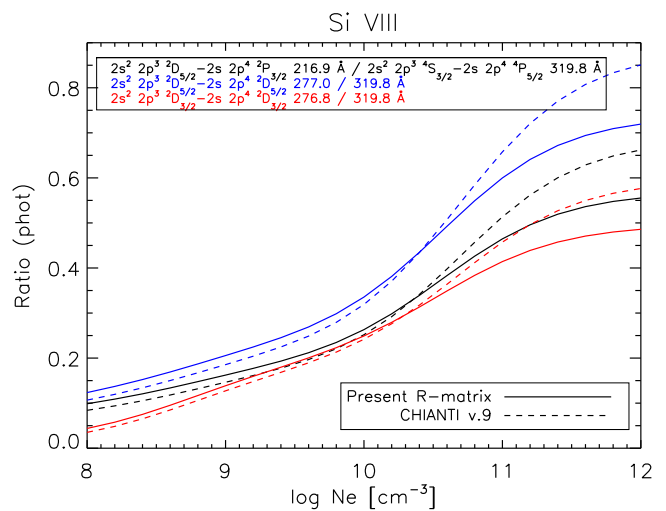}
\caption{Line ratios (in ${\rm phot~cm^{-2}~s^{-1}}$) of selected resonance lines (Table~\ref{tbl:gml_tbl1}) of \ion{Si}{VIII} as a function of density. The solid curves are calculated with the present $R$-matrix data, while the dashed curves use the $R$-matrix data of \citet{bel01} as incorporated in CHIANTI v9.0.1. }
\label{fig:si_8_ratios}
\end{figure}

Within the $2s^22p^3$ ground configuration, the two forbidden transitions at 1440.5~\AA\ and 1445.7~\AA\ (Table~\ref{tbl:gml_tbl1}), are the most important plasma diagnostic lines. As there is little difference between the $R$-matrix data of \citet{bel01} and the present work (Fig.~\ref{fig:plot_ecs_071408}) at $T\sim10^6$~K, for solar observations, the electron density derived from the the line ratio of the two agree well ($\sim3$~\%). Several other density diagnostic line ratios are also available in the EUV band. Three of them are displayed in Fig.~\ref{fig:si_8_ratios}. For these lines, some differences between the previous CHIANTI model and the present one are clear, especially at higher densities. A detailed comparison with solar observations is complicated by the fact that the 276.8~\AA\ and 277.0~\AA\ lines are blended with transitions from other ions, and is therefore beyond the scope of this paper. 

\subsection{\ion{O}{II}}
\label{sct:070802}
The most recent $R$-matrix calculations of electron-impact excitation data for \ion{O}{II} (or ${\rm O^{+}}$) are presented in \citet[][T07]{tay07} and \citet[][K09]{kis09}.

The present work and T07 used AUTOSTRUCTURE and multi-configuration Hartree-Fock for the atomic structure calculations, respectively. As shown in the bottom-right panel of Fig.~\ref{fig:plot_cflev}, the level energies of T07 agree better with respect to NIST than the present work, especially for the low-lying transitions. The transition strengths of NIST, T07, and the present work agree well with each other for $\log (gf)\gtrsim-4$ (the bottom-left panel of Fig.~\ref{fig:plot_cftran}). 

The present work, T07, and K09 used the $R$-matrix ICFT method, B-spline $R$-matrix method, and Breit–Pauli $R$-matrix method with pseudo-states for the scattering calculations, respectively. T07 provided effective collision strengths for transitions between the lowest 47 energy levels, while K09 focused on transitions between the lowest five energy levels. 

\begin{figure*}
\centering
\includegraphics[width=.8\hsize, trim={0.5cm 0.cm 1.5cm 0.5cm}, clip]{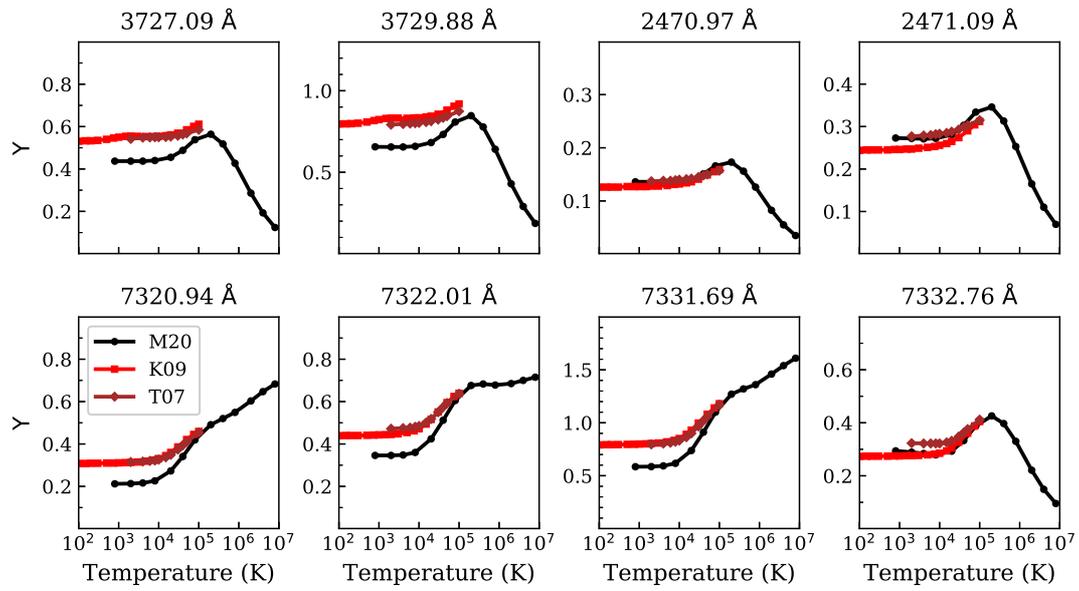}
\caption{Comparison of \ion{O}{II} (or ${\rm O^{+}}$) effective collision strengths between the present work (M20), \citet[][T07]{tay07}, and \citet[][K09]{kis09} for eight common transitions from the ground and metastable levels. }
\label{fig:plot_ecs_070802}
\end{figure*}

In Fig.~\ref{fig:plot_ecs_070802}, we compare effective collision strengths of \ion{O}{II} (or ${\rm O^{+}}$) between the present work (M20), \citet[][T07]{tay07}, and \citet[][K09]{kis09} for eight common transitions from the ground and metastable levels (Table~\ref{tbl:gml_tbl2}). Both T07 and K09 agree with each other for effective collision strengths at lower temperatures ($T\lesssim10^5$~K) and, thus, are recommended. We consider our present work less accurate at lower temperatures, mainly due to the poorer atomic structure (Fig.~\ref{fig:plot_cflev} and Fig.~\ref{fig:plot_cftran}). In general, calculations with non-orthogonal orbitals and pseudo states should be preferred for low-charge ions like \ion{O}{II}, as in our case we are limited by the use of  orthogonal orbitals for our $R$-matrix calculations (Section~\ref{sct:str}). For transitions not covered by T07 and K09, caution should be exercised  when incorporating our data for lower temperatures into atomic databases, e.g., as used by photoionization plasma codes, due to its relatively low accuracy. On the other hand, the present work agrees better with T07 and K09 at $T\gtrsim10^5$~K. Thus, the effective collision strengths at higher temperatures ($T\gtrsim10^5$~K), typical of collisional plasmas, can be incorporated freely into the atomic databases of plasma codes.

\section{Conclusion}
\label{sct:sum}
We have presented a systematic set of $R$-matrix intermediate-coupling frame transfer calculations for N-like ions from \ion{O}{II} to \ion{Zn}{XXIV} (i.e., O$^{+}$ to Zn$^{23+}$) to obtain level-resolved effective collision strengths over a wide temperature range. The extensive comparison made with results in the literature for a sample of ions important for astrophysical applications provides a reassuring picture. Our effective collision strengths from the ground and metastable levels agree, in general, within 0.2 dex with previous state-of-the-art calculations, at temperatures relevant to modelling. Our configuration interaction target and close-coupling collision expansion are significantly larger than previous studies. This indicates that we have reached convergence here. On the other hand, as we have seen in previous studies, collision strengths involving the highest-lying energy levels are not converged.

As accurate $R$-matrix data were available for only some ions, the present calculations are a significant extension and improvement for this iso-electronic sequence. For several minor (cosmicly rare) ions such as \ion{Ti}{XV} and \ion{Cr}{XVII}, the present data are a significant improvement with respect to previous distorted-wave calculations.

We have shown examples where significant differences 
are found in the diagnostics (densities, abundances) when compared to CHIANTI models which used only distorted wave data (Ca {\sc xiv} and Ar {\sc xii}). Some differences are present also when previous $R$-matrix data are utilized (Si {\sc viii}).

The effective collision strengths are archived according to the Atomic Data and Analysis Structure (ADAS) data class {\it adf04} and will be available in OPEN-ADAS and our UK-APAP website. These data will be incorporated into plasma codes like CHIANTI \citep{der97,der19} and SPEX \citep{kaa96,kaa18}. These data can improve the quality of plasma diagnostics especially in the context of future high-resolution spectrometers. We plan to perform similar calculations for the O-like iso-electronic sequence. 

\begin{acknowledgements}
      The present work is funded by STFC (UK) through the University of Strathclyde UK APAP network grant ST/R000743/1 and the University of Cambridge DAMTP atomic astrophysics group grants ST/P000665/1 and ST/T000481/1. We thank Enrico Landi for providing the observed intensities of the SUMER lines. JM thanks useful discussions with Helen Mason, Martin O Mullane, and Pete Storey. JM acknowledges atomic data provided by G. Jiang. We thank the referee for careful reading of the manuscript and useful suggestions. 
\end{acknowledgements}




\appendix
\section{\ion{Ar}{XII}}
\label{sct:071812app}
\citet{lud10} performed an ICFT $R$-matrix electron-impact excitation calculation for \ion{Ar}{XII}. They included 186 fine-structure levels of the target ion. The effective collision strengths are available over a wide temperature range (between $2.88\times10^4$~K and $2.88\times10^7$~K). 

For transitions involving levels \#158 to \#186 (their highest energy level), the effective collision strengths at the lowest temperature ($2.88\times10^4$~K) are either zero or $\gtrsim5$ orders of magnitude smaller than that of the next temperature point ($7.20\times10^5$~K). A similar jump with $\gtrsim4$ orders of magnitude is also found between effective collision strengths at $7.20\times10^4$~K and $1.44\times10^5$~K for most of the transitions involving levels \#158 to \#186. 

\section{\ion{Mg}{VI}}
\label{sct:071206}
The most recent $R$-matrix calculation of electron-impact excitation data for \ion{Mg}{VI} (or ${\rm Mg^{5+}}$) is presented by \citet[][W18]{wan18}. In addition, a data set provided by \citet[][W05]{wit05} is also available from OPEN-ADAS without an associated publication.


W05, W18 and the present work all used the ICFT $R$-matrix method for the scattering calculation. Both W05 and W18 included 272 fine-structure levels of the target ion. For W18, effective collision strengths from the ground level to the lowest 120 levels are tabulated in their Table 22 for \ion{Mg}{VI}. For W05, according to the comments in the adf04 file, the atomic structure is optimized for transitions within $n=2$ only. 

\begin{figure}
\centering
\includegraphics[width=.87\hsize, trim={0.cm 0.5cm 0.5cm 0.5cm}, clip]{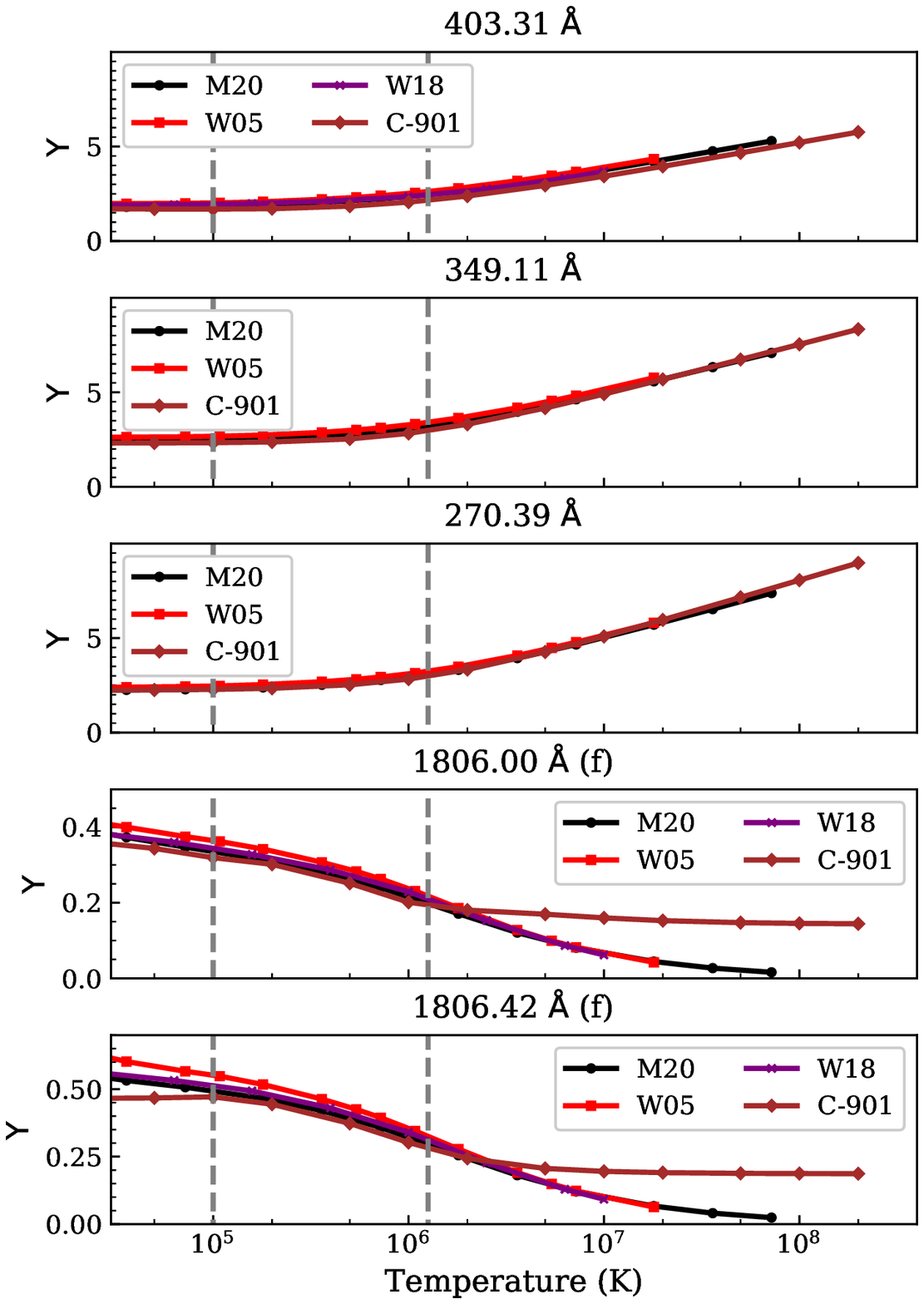}
\caption{Comparison of \ion{Mg}{VI} (or ${\rm Mg^{5+}}$) effective collision strengths between the present work (M20), \citet[][W05]{wit05}, \citet[][W18]{wan18} and \citet[][$R$-matrix]{ram97} as incorporated in the CHIANTI atomic database v9.0.1 (C-901) for selected transitions listed in Table~\ref{tbl:gml_tbl2}. The top panel is a dipole transition from the ground level, followed by two metastable transitions. The bottom two panels are forbidden transitions from the ground level to the first two metastable levels. The vertical dashed lines indicate the temperature range originally provided by \citet{ram97}. The brown diamonds outside this temperature range is extrapolated in CHIANTI. }
\label{fig:plot_ecs_071206}
\end{figure}

In Fig.~\ref{fig:plot_ecs_071206}, we compare the effective collision strengths of selected three transitions from the ground and metastable levels listed in Table~\ref{tbl:gml_tbl2}. For the three dipole transitions from the ground and metastable levels, we found good agreement between the three data sets: present work (M20), \citet[][W05]{wit05}, and \citet[][$R$-matrix]{ram97} as incorporated in the CHIANTI atomic database v9.0.1 (C-901). For the two forbidden transitions from the ground to the first two metastable levels, the extrapolation at higher temperatures in the current version of CHIANTI is inaccurate.  

\begin{table}[!h]
\caption[]{Selected prominent transitions from the lowest three energy levels for \ion{Mg}{VI}. The rest-frame wavelength (\AA) are taken from the CHIANTI atomic database. }
\label{tbl:gml_tbl2}
\centering
\begin{tabular}{lll}
\hline\hline
\noalign{\smallskip} 
Lower level & Upper level & $\lambda_0$ ($\AA$) \\
\noalign{\smallskip} 
\hline
\noalign{\smallskip} 
$2s^2 2p^3~(^4S_{3/2})$ & $2s 2p^4~(^2P_{5/2})$ & 403.01 \\
$2s^2 2p^3~(^2D_{5/2})$ & $2s 2p^4~(^2D_{3/2})$ & 349.11 \\
$2s^2 2p^3~(^2D_{5/2})$ & $2s 2p^4~(^2P_{3/2})$ & 270.39 \\
$2s^2 2p^3~(^4S_{3/2})$ & $2s^2 2p^3~(^2D_{3/2})$ & 1806.00 (f) \\
$2s^2 2p^3~(^4S_{3/2})$ & $2s^2 2p^3~(^2D_{5/2})$ & 1806.42 (f) \\
\noalign{\smallskip} 
\hline
\end{tabular}
\end{table}

 \end{document}